\documentclass[aps,prd,
twocolumn,
 amsmath,amssymb,
superscriptaddress]{revtex4-1}

\usepackage{graphicx}
\usepackage{dcolumn}
\usepackage{bm}
\usepackage{epstopdf}

\usepackage[mathlines]{lineno}

\begin{document}

\preprint{PRD...}

\title{Testing Non-Standard Interactions Between Solar Neutrinos and Quarks with Super-Kamiokande}

\newcommand{\AFFicrr}{\affiliation{Kamioka Observatory, Institute for Cosmic Ray Research, University of Tokyo, Kamioka, Gifu 506-1205, Japan}}
\newcommand{\AFFkashiwa}{\affiliation{Research Center for Cosmic Neutrinos, Institute for Cosmic Ray Research, University of Tokyo, Kashiwa, Chiba 277-8582, Japan}}
\newcommand{\AFFicrrkashiwa}{\affiliation{Institute for Cosmic Ray Research, University of Tokyo, Kashiwa, Chiba 277-8582, Japan}}
\newcommand{\AFFipmu}{\affiliation{Kavli Institute for the Physics and
Mathematics of the Universe (WPI), The University of Tokyo Institutes for Advanced Study,
University of Tokyo, Kashiwa, Chiba 277-8583, Japan }}
\newcommand{\AFFmad}{\affiliation{Department of Theoretical Physics, University Autonoma Madrid, 28049 Madrid, Spain}}
\newcommand{\AFFubc}{\affiliation{Department of Physics and Astronomy, University of British Columbia, Vancouver, BC, V6T1Z4, Canada}}
\newcommand{\AFFbu}{\affiliation{Department of Physics, Boston University, Boston, MA 02215, USA}}
\newcommand{\AFFbci}{\affiliation{Department of Physics, British Columbia Institute of Technology, Burnaby, BC, V5G 3H2, Canada}}
\newcommand{\AFFuci}{\affiliation{Department of Physics and Astronomy, University of California, Irvine, Irvine, CA 92697-4575, USA }}
\newcommand{\AFFcsu}{\affiliation{Department of Physics, California State University, Dominguez Hills, Carson, CA 90747, USA}}
\newcommand{\AFFcnm}{\affiliation{Department of Physics, Chonnam National University, Kwangju 500-757, Korea}}
\newcommand{\AFFduke}{\affiliation{Department of Physics, Duke University, Durham NC 27708, USA}}
\newcommand{\AFFfukuoka}{\affiliation{Junior College, Fukuoka Institute of Technology, Fukuoka, Fukuoka 811-0295, Japan}}
\newcommand{\AFFgifu}{\affiliation{Department of Physics, Gifu University, Gifu, Gifu 501-1193, Japan}}
\newcommand{\AFFgist}{\affiliation{GIST College, Gwangju Institute of Science and Technology, Gwangju 500-712, Korea}}
\newcommand{\AFFuh}{\affiliation{Department of Physics and Astronomy, University of Hawaii, Honolulu, HI 96822, USA}}
\newcommand{\AFFicl}{\affiliation{Department of Physics, Imperial College London , London, SW7 2AZ, United Kingdom }}
\newcommand{\AFFkeio}{\affiliation{Department of Physics, Keio University, Yokohama, Kanagawa, 223-8522, Japan}}
\newcommand{\AFFkek}{\affiliation{High Energy Accelerator Research Organization (KEK), Tsukuba, Ibaraki 305-0801, Japan }}
\newcommand{\AFFkcl}{\affiliation{Department of Physics, King’s College London, London, WC2R 2LS, UK}}
\newcommand{\AFFkobe}{\affiliation{Department of Physics, Kobe University, Kobe, Hyogo 657-8501, Japan}}
\newcommand{\AFFkyoto}{\affiliation{Department of Physics, Kyoto University, Kyoto, Kyoto 606-8502, Japan}}
\newcommand{\AFFliv}{\affiliation{Department of Physics, University of Liverpool, Liverpool, L69 7ZE, United Kingdom}}
\newcommand{\AFFmiyagi}{\affiliation{Department of Physics, Miyagi University of Education, Sendai, Miyagi 980-0845, Japan}}
\newcommand{\AFFnagoya}{\affiliation{Institute for Space-Earth Environmental Research, Nagoya University, Nagoya, Aichi 464-8602, Japan}}
\newcommand{\AFFkmi}{\affiliation{Kobayashi-Maskawa Institute for the Origin of Particles and the Universe, Nagoya University, Nagoya, Aichi 464-8602, Japan}}
\newcommand{\AFFpol}{\affiliation{National Centre For Nuclear Research, 00-681 Warsaw, Poland}}
\newcommand{\AFFwarsaw}{\affiliation{Faculty of Physics, University of Warsaw, Warsaw, 02-093, Poland}}
\newcommand{\AFFsuny}{\affiliation{Department of Physics and Astronomy, State University of New York at Stony Brook, NY 11794-3800, USA}}
\newcommand{\AFFokayama}{\affiliation{Department of Physics, Okayama University, Okayama, Okayama 700-8530, Japan }}
\newcommand{\AFFosaka}{\affiliation{Department of Physics, Osaka University, Toyonaka, Osaka 560-0043, Japan}}
\newcommand{\AFFox}{\affiliation{Department of Physics, Oxford University, Oxford, OX1 3PU, United Kingdom}}
\newcommand{\AFFqmul}{\affiliation{School of Physics and Astronomy, Queen Mary University of London, London, E1 4NS, United Kingdom}}
\newcommand{\AFFregina}{\affiliation{Department of Physics, University of Regina, 3737 Wascana Parkway, Regina, SK, S4SOA2, Canada}}
\newcommand{\AFFseoul}{\affiliation{Department of Physics, Seoul National University, Seoul 151-742, Korea}}
\newcommand{\AFFsheff}{\affiliation{Department of Physics and Astronomy, University of Sheffield, S3 7RH, Sheffield, United Kingdom}}
\newcommand{\AFFshizuokasc}{\affiliation{Department of Informatics in
Social Welfare, Shizuoka University of Welfare, Yaizu, Shizuoka, 425-8611, Japan}}
\newcommand{\AFFstfc}{\affiliation{STFC, Rutherford Appleton Laboratory, Harwell Oxford, and Daresbury Laboratory, Warrington, OX11 0QX, United Kingdom}}
\newcommand{\AFFskk}{\affiliation{Department of Physics, Sungkyunkwan University, Suwon 440-746, Korea}}
\newcommand{\AFFtokyo}{\affiliation{The University of Tokyo, Bunkyo, Tokyo 113-0033, Japan }}
\newcommand{\AFFtodai}{\affiliation{Department of Physics, University of Tokyo, Bunkyo, Tokyo 113-0033, Japan }}
\newcommand{\AFFtit}{\affiliation{Department of Physics,Tokyo Institute of Technology, Meguro, Tokyo 152-8551, Japan }}
\newcommand{\AFFtus}{\affiliation{Department of Physics, Faculty of Science and Technology, Tokyo University of Science, Noda, Chiba 278-8510, Japan }}
\newcommand{\AFFtoronto}{\affiliation{Department of Physics, University of Toronto, ON, M5S 1A7, Canada }}
\newcommand{\AFFtriumf}{\affiliation{TRIUMF, 4004 Wesbrook Mall, Vancouver, BC, V6T2A3, Canada }}
\newcommand{\AFFtokai}{\affiliation{Department of Physics, Tokai University, Hiratsuka, Kanagawa 259-1292, Japan}}
\newcommand{\AFFtohoku}{\affiliation{Department of Physics, Faculty of Science, Tohoku University, Sendai, Miyagi, 980-8578, Japan}}
\newcommand{\AFFtsinghua}{\affiliation{Department of Engineering Physics, Tsinghua University, Beijing, 100084, China}}
\newcommand{\AFFynu}{\affiliation{Faculty of Engineering, Yokohama National University, Yokohama, 240-8501, Japan}}
\newcommand{\AFFuwarwick}{\affiliation{Department of Physics, University of Warwick, Coventry, CV4 7AL, UK}}
\newcommand{\AFFuwinnipeg}{\affiliation{Department of Physics, University of Winnipeg, MB R3J 3L8, Canada}}
\newcommand{\AFFllr}{\affiliation{Ecole Polytechnique, IN2P3-CNRS, Laboratoire Leprince-Ringuet, F-91120 Palaiseau, France }}
\newcommand{\AFFbari}{\affiliation{ Dipartimento Interuniversitario di Fisica, INFN Sezione di Bari and Universit\`a e Politecnico di Bari, I-70125, Bari, Italy}}
\newcommand{\AFFnapoli}{\affiliation{Dipartimento di Fisica, INFN Sezione di Napoli and Universit\`a di Napoli, I-80126, Napoli, Italy}}
\newcommand{\AFFroma}{\affiliation{INFN Sezione di Roma and Universit\`a di Roma ``La Sapienza'', I-00185, Roma, Italy}}
\newcommand{\AFFpadova}{\affiliation{Dipartimento di Fisica, INFN Sezione di Padova and Universit\`a di Padova, I-35131, Padova, Italy}}
\newcommand{\AFFilance}{\affiliation{International Laboratory for Astrophysics, Neutrino and Cosmology Experiment, Kashiwa, Chiba 277-8582, Japan}}
\newcommand{\AFFifirse}{\affiliation{Institute For Interdisciplinary Research in Science and Education, ICISE, Quy Nhon, 55121, Vietnam}}

\AFFicrr
\AFFkashiwa
\AFFicrrkashiwa
\AFFmad
\AFFbu
\AFFbci
\AFFubc
\AFFuci
\AFFcsu
\AFFcnm
\AFFduke
\AFFllr
\AFFfukuoka
\AFFgifu
\AFFgist
\AFFuh
\AFFifirse
\AFFicl
\AFFbari
\AFFnapoli
\AFFpadova
\AFFroma
\AFFilance
\AFFkeio
\AFFkek
\AFFkcl
\AFFkobe
\AFFkyoto
\AFFliv
\AFFmiyagi
\AFFnagoya
\AFFkmi
\AFFpol
\AFFsuny
\AFFokayama
\AFFosaka
\AFFox
\AFFqmul
\AFFregina
\AFFseoul
\AFFsheff
\AFFshizuokasc
\AFFstfc
\AFFskk
\AFFtohoku
\AFFtokai
\AFFtokyo
\AFFtodai
\AFFipmu
\AFFtit
\AFFtus
\AFFtoronto
\AFFtriumf
\AFFtsinghua
\AFFwarsaw
\AFFuwarwick
\AFFuwinnipeg
\AFFynu

\author{P.~Weatherly} 
\AFFuci

\author{K.~Abe}
\AFFicrr
\AFFipmu
\author{C.~Bronner}
\AFFicrr
\author{Y.~Hayato}
\AFFicrr
\AFFipmu
\author{K.~Hiraide}
\AFFicrr
\AFFipmu
\author{M.~Ikeda}
\AFFicrr
\AFFipmu

\author{K.~Iyogi}
\AFFicrr 
\author{J.~Kameda}
\AFFicrr
\AFFipmu 
\author{Y.~Kanemura}
\AFFicrr
\author{Y.~Kataoka}
\AFFicrr
\AFFipmu 
% ...


\collaboration{The Super-Kamiokande Collaboration}
\noaffiliation

\date{\today}

\begin{abstract}
Non-Standard Interactions (NSI) between neutrinos and matter affect the neutrino flavor oscillations.
Due to the high matter density in the core of the Sun, solar neutrinos are 
 suited to probe these interactions.
Using the $277$ kton-yr exposure of Super-Kamiokande to $^{8}$B solar neutrinos, we search for the presence of NSI.
Our data favors the presence of NSI with down quarks at 1.8$\sigma$, and with up quarks at 1.6$\sigma$, with the best fit NSI parameters being ($\epsilon_{11}^{d},\epsilon_{12}^{d}$) = (-3.3, -3.1) for $d$-quarks and 
($\epsilon_{11}^{u},\epsilon_{12}^{u}$) = (-2.5, -3.1) for $u$-quarks.
After combining with data from the Sudbury Neutrino Observatory and Borexino, the significance increases by 0.1$\sigma$.
\end{abstract}

\pacs{Valid PACS appear here}
\maketitle

\section{
  \label{sec:intro}
  Introduction
}
Current flux measurements of the solar neutrinos produced in the nuclear $pp$ fusion chain in the solar core indicate that there is an energy-dependent transition in the three-flavor electron neutrino survival probability $(P^{3f}_{ee}$, or simply $P_{ee})$.
The transition is from a higher probability 
at low neutrino energy, $E_\nu \lesssim$ 1 MeV, to a distinctly lower probability 
 above $\sim$5 MeV.
The low energy $P_{ee}$ is consistent with solar neutrinos whose flavor content are dominated by vacuum oscillations parameterized by the Pontecorvo--Maki-Nakagawa-Sakata (PMNS) \cite{P_paper,*MNS_paper}, matrix and squared-mass splittings, and whose fluxes are predicted by Standard Solar Models (SSMs) \cite{bethe_ppfusion, *cno_1_vW, *cno_2_vW, *kz_ssm, *haxton_solar_nu_problem}\cite{bahcall_ssm_64, *bahcall_ssm_88, *bahcall_ssm_92, *BPB2001, *bahcall_web}\cite{gs98, *agss09}\cite{bp04}\cite{bp2000}\cite{bs05}\cite{new_ssms}.
At higher energies, the suppression of the $P_{ee}$ is consistent with the SSMs and the PMNS and squared-mass splitting parameterization of neutrino oscillations, if the flavor conversion of the higher energy $^{8}$B and $hep$ solar neutrinos undergo adiabatic conversion within the Sun. 
This flavor conversion of the higher energy solar neutrinos is due to the high matter density in the solar core and is predicted by the Mikheyev-Smirnov-Wolfenstein (MSW) effect  \cite{mswpaper,*mswpaper0}.  
The solar neutrinos produced by the Sun offer a unique window into determining the effects matter can have on neutrino oscillations since: a) neutrinos are produced only in a pure $\nu_{e}$ flavor eigenstate, b) they are produced a high rate, c) they are produced in a region of high matter density with large path-lengths, and d) they are produced with varying energy spectra due to the underlying parent processes in $pp$ fusion.

Borexino's simultaneous fit to all the $pp$-chain solar neutrinos \cite{borexino_nature, *borexino_cno_Nature} 
resulted in a measurement of the low energy $pp$ solar neutrino flux that indicates a $P_{ee}$ value around $\sim55\%$.
The flux measurements of $^{8}$B and $hep$ solar neutrinos by Super-Kamiokande (SK) \cite{SK4Solar} and the Sudbury Nutrino Observatory (SNO) \cite{snob8paper,*snoheppaper,*sno3phase} both constrain the electron survival probability of the solar neutrinos arriving during the day $P^{\textrm{DAY}}_{ee}$ to be around $33\%$, providing evidence for MSW-induced flavor conversion 
of the electron neutrinos produced in the core of the Sun into muon or tau flavored neutrinos. 
The data from the SK recoil electron spectra, as well as the published SNO results of their polynomial fit to the solar neutrino spectrum, favor a flatter recoil electron spectrum and a flatter $P_{ee}$ distribution than expected. 

Including the solar neutrino flux measurements by the radiochemical experiments Homestake \cite{homestake,*homestake_2}, SAGE \cite{sage}, and GALLEX/GNO \cite{gallex,*gno}, and also the recent measurements of the $^8$B flux by SNO$+$ \cite{sno_plus}, all solar neutrino data all are consistent with the differing $P_{ee}$ between the low energy and high energy solar neutrinos, implying a transition between the two regions.
This transition is sometimes referred to as the ``upturn" between the higher energy neutrinos with lower $P_{ee}$ to the lower energy neutrinos with higher $P_{ee}$.
Independent measurements by KamLAND of flux for the $pp$-chain 862 keV $^{7}\textrm{Be}$ solar neutrinos \cite{kamland_be7} and for the higher energy $^8$B solar neutrino flux \cite{kamland_b8} are consistent with this picture. 
The best fit region for all solar neutrino experiments in the 
 ($\sin^{2}\theta_{12},\Delta m_{21}^{2})$ parameter space is the so-called ``Large Mixing Angle'' (LMA) region.
The LMA region lies between $0.12 < \sin^{2}\theta_{12} < 0.45$ at the 7.5$\sigma$ level and excludes values of $\Delta m_{21}^{2}$ below (above) $1.33 \times 10^{-5}$ $(1.9\times10^{-4})$  eV$^{2}$ at 5.5$\sigma$ (7.5$\sigma$) \cite{SK4Solar}.
Additionally, data from the reactor neutrino experiments Daya Bay \cite{daya_bay_2017,*daya_bay_hydrogen,*dayabay}, RENO \cite{reno_2017, *reno}, and Double Chooz \cite{doublechooz_2014, *doublechooz} strongly constrain the value of  $\sin^{2}\theta_{13} =  0.0219 \pm 0.0014$, where the uncertainty is at the 1$\sigma$ level.
The most recent world average determined by the Particle Data Group is $\sin^{2}\theta_{13} =  0.0218 \pm 0.0007$ \cite{pdg_theta13}.

Additional effects on solar neutrino oscillation arise due to solar neutrinos passing through the Earth during the night. 
As solar neutrinos pass through the Earth's corresponding matter potential, the additional effect on their wave function can lead to enhancements in their oscillations, including regeneration into $\nu_{e}$ states that are not seen for neutrinos arriving during the daytime \cite{petcov_98,  *baltz_weneser_87, *carlson_86, *msw_regen_in_earth_86}.
The day-night asymmetry in the rate of events detected during the day $(r_{D})$ and night $(r_{N})$ is given by 
\begin{equation}
\label{adn_eq}
A_\textrm{DN} = \frac{r_\textrm{D} - r_\textrm{N}}{\frac{1}{2}(r_\textrm{D} + r_\textrm{N})},
\end{equation}
and is another testable signature for neutrino oscillations.
The $A_\textrm{DN}$ arises in the detection rate of both neutrino-electron elastic scattering (ES) and charged current (CC) interactions on D$_{2}$O.  
In the LMA region, the matter effects in the Earth causes an enhancement of the $\nu_{e}$ component of the neutrino wavefunction, leading to a negative day-night asymmetry. 
The combined phases of SK give a day-night asymmetry measurement of 
$A_\textrm{DN} = (-4.2 \pm 1.2(\textrm{stat.}) \pm  0.8(\textrm{syst.}) )\%$ \cite{SK4Solar}
when using Eq.\ (\ref{adn_eq}), and, by using the day-night amplitude fit method described in \cite{dnpaper}, a measured asymmetry of $A_\textrm{DN}^{\textrm{fit}} = (-3.2 \pm 1.1(\textrm{stat.})\pm 0.5(\textrm{syst.}))\%$.

As the SK spectral data is not favoring 
the upturn, other phenomena distorting 
the shape of the $P_{ee}$ distribution could be tested to determine how consistent those models are with what is measured by SK (and SNO), as well as determining how strong those interactions could be. 
Non-Standard Interactions (NSI) between solar neutrinos and matter in the Sun and Earth is one such model that 
affects neutrino oscillation.
NSI modifies the $P_{ee}$ in various ways, from delaying the onset of the transition to adiabatic conversion, 
 to flattening the low energy upturn, or even inverting it into a downturn, prior to returning to the vacuum oscillation probability levels at low neutrino energy \cite{Friedland04}\cite{3flavNSI2013}. 
Additionally, for strong NSI, matter effects in the Earth become more important, and can also induce regeneration of the $\nu_{e}$ flavor state in the lower energy region of the spectrum than is the case with Standard Interactions (SI), i.e. no NSI present.
It should be noted here that the PMNS and squared-mass splitting parameterization for neutrino oscillations in vacuum and effects arising from neutrino-electron interactions in matter are referred to in this work as standard interactions for neutrinos.

External fits for NSI parameters by theorists have been performed using publicly available SK data \cite{Friedland04}\cite{3flavNSI2013}.
However, these fits are unable to incorporate the SK detector simulations, systematic uncertainties, correlations, recoil electron spectral and $A_\textrm{DN}$ predictions, and other fine details of the SK solar neutrino oscillation analysis only the SK collaboration can provide, including our combined fit with SNO.
An analysis of the effective NSI parameters describing the strength of the interactions between quarks and neutrinos using the tools of the SK solar neutrino oscillation analysis is presented in this work.
Details of these effective parameters are provided in Sec. \ref{sec:nsi}.

These NSI terms only affect the $P_{ee}$ in the MSW sensitive region of $E_{\nu}/\Delta m_{21}^{2}$ and can cause distortions to the $P_{ee}$ when transitioning into and out of this region. 
However, the contribution from experiments sensitive to the lower energy solar neutrinos to the constraints on the NSI parameters tested in this analysis
 is minimal compared to SK and SNO, due to the lack of matter effects at lower energies \cite{3flavNSI2013}.
Stronger constraints are made possible by additionally fitting to KamLAND reactor $\bar{\nu}_{e}$ data \cite{kamland11,*kamland08} after including the NSI effects in the Earth matter potential,  as in \cite{3flavNSI2013}.
This is because KamLAND tightly constrains the value of $\Delta m_{21}^{2}$ more than the solar data alone and is mostly insensitive to NSI in the matter potential \cite{3flavNSI2013}.   
In the case of Borexino, while their sensitivity to the higher energy $^{8}$B neutrinos is not as good as SK and SNO, 
their measurement of a very small asymmetry between the rate of 862 keV $^{7}$Be neutrinos during the day versus the night ($A_\textrm{DN}^{\textrm{Be7}}$) rejects the best-fit point in the so-called ``LOW'' region at more than 8.5$\sigma$, a stronger rejection than SK \cite{borexino_be7_adn}.
The LOW region has a similar $\sin^{2}\theta_{12}$ range as the LMA region, but lives in a smaller $\Delta m_{21}^{2}$ region between $10^{-8} \textrm{eV}^{2}$ and $10^{-6} \textrm{eV}^{2}$.
Borexino's $A_\textrm{DN}^{\textrm{Be7}}$ measurement additionally disfavors the LOW region at greater than 3$\sigma$. 
This, along with Borexino's strong rejection of the best-fit point in the LOW region, is useful to help further constrain the NSI parameters tested in this analysis.
It should be noted that the KamLAND measurement of the  862 keV $^{7}$Be solar neutrino flux is not included in this analysis since they do not report a day-night asymmetry in their work.

\subsection{\label{sec:sk_intro}The Super-Kamiokande Detector}
The SK detector \cite{sknim} is a large water Cherenkov detector located one kilometer below the peak of Mount Ikenoyama, outside of Kamioka Town in the Gifu prefecture of Japan. The detector is an upright cylinder containing 50 kilotons of ultrapure water, with a 32.5 kton cylindrical inner detector (ID) and an outer detector (OD) 2.5 m thick. 
The ID and OD are optically separated, and about 2 m of the OD is used as an active veto. The half-meter thick structure that separates the ID from the OD houses the photomultiplier tubes (PMTs) used to collect the emitted light from charged particles traveling above the Cherenkov threshold in the water. 
Further details of the SK detector can be found in \cite{sknim} and \cite{sk4_calib_nim_14}.

Solar neutrinos interact in SK via elastic scattering (ES) off of electrons in the water. 
The Cherenkov light produced by these recoil electrons is collected by 50 cm (20 inch) PMTs in the SK ID.
The interaction vertex and direction the recoil electron travels can be reconstructed using the timing and position of the hit PMTs. 
Additionally, the number of hit PMTs is directly correlated to the kinetic energy of the recoil electron.
While this electroweak process has no energy threshold, these recoil electrons can only be seen if they are above Cherenkov threshold and produce enough photons for the reconstruction algorithms to piece the information back together.
Because the direction of the recoil electron is strongly forward-peaked relative to the incident neutrino and because the position of the Sun is well known at any given time, a solar signal can be statistically extracted based off of the cosine of the solar angle, $\cos\theta_\textrm{Sun}$. 
The majority of the solar neutrino signal dominates over the background for $\cos\theta_\textrm{Sun} > 0.8$ \cite{SK4Solar}. 

The various backgrounds in SK's solar neutrino analysis are rejected by imposing a series of requirements.
Calibration runs and runs with hardware or software trouble are removed from the analysis.
Quality requirements are imposed for vertex reconstruction based on the timing goodness of the PMT timing residuals and the azimuthal symmetry of PMT hit pattern made by the Cherenkov cone to remove radioactive backgrounds originating from the PMT structure, glass, and enclosures. 
Cutting events with small hit clusters and events pointing into the detector that originate close to the PMT wall (the so-called external event cut)  helps to further remove this class of background.
Consistency in the PMT hit pattern with the $42^{\circ}$ opening angle of the Cherenkov cone removes events with multiple cones like $\beta$-decay of isotopes to excited nuclear states that emit de-excitation $\gamma$s. 
Events below 4.99 MeV reconstructed around known calibration sources and cables that have some radioactive contaminants are removed.
Cosmic-ray $\mu$ events (number hit PMTs $>$ 400 $\simeq$ 60 MeV for recoil $e$-like events) are rejected. 
Sometimes, these $\mu$ can break up an oxygen nucleus, causing long-lived radioactive isotopes called spallation events \cite{li_beacom_spall_2014, *li_beacom_spall_2015}, which are rejected using a spallation likelihood function \cite{SK1Solar}\cite{yzhang_spallation_paper, *kirk_srn}. 
For $^{16}$N produced when lower energy cosmic-ray $\mu$ capture on $^{16}$O and which decay to electrons and/or gamma-rays, a series of space-time correlation cuts are applied between the $\mu$ and $e/\gamma$s to remove these backgrounds \cite{n16_as_sk_calib_source}.
Finally, a fiducialization cut requires events occurring within 2 m the PMT wall to be rejected, leaving a fiducial volume of 22.5 kton. Below 4.99 MeV, tighter, more complicated constraints are implemented as backgrounds ingress further to the center of the detector. 
For more details on the SK-IV solar neutrino event selection criteria, refer to Sec. III A in \cite{SK4Solar}.

These quality requirements also remove backgrounds resulting from radioactive radon gas present in the air of the mine.
The radon makes its way into SK through the water purification system, and is concentrated at the system's injection sites at the bottom of the detector.
The temperature of the detector is finely controlled to stop convection of the water in SK and produce a bottom-to-top laminar flow of the water.
This technique allows these isotopes decay away in the detector's bottom region which helps maximize the fiducial volume.
Measurements of radon concentrations from various locations throughout the detector and water purification system during SK-IV have been reported by Ref.\ \cite{nakano_radon_paper}.

SK detects approximately 18 of the 280 solar neutrino interactions occurring within the fiducial volume per day, 
with an exposure spanning twenty years. 
In the analysis presented in this paper, 
NSI parameters relevant to solar neutrinos will be probed using the recoil electron spectrum data from SK, taken up until the beginning of February 2014 (denoted as SK-Only). 
This data includes the 1496 days of live time of SK-I data \cite{SK1Solar}, the 792 days of SK-II \cite{SK2Solar}, the 549 days of SK-III \cite{SK3Solar}, and the 1664 days from the first half of SK-IV \cite{SK4Solar}.
With these statistics and the detector's energy resolution, the recoil electron spectrum is binned to one MeV between 12.5 MeV and 15.5 MeV and, to half an MeV below 12.5 MeV with the energy threshold of the solar neutrino spectral analysis differing for each experimental phase: 4.49 MeV for SK-I, 6.49 MeV for SK-II, 3.99 MeV for SK-III, and 3.49 MeV for SK-IV.
This allows for an accurate measurement of $^{8}$B neutrino flux with a precise measure of the $^{8}$B recoil electron spectrum during both the day and the night, and enables SK the ability to study matter effects on  $^{8}$B solar neutrinos from both matter in the Sun and in the Earth.
The recoil electron spectra for each phase of SK are provided in Table \ref{recoil_e_spec_table} in units of the measured number of events in SK divided by the expected number of events predicted by the BP2004 SSM \cite{bp04} assuming no neutrino oscillations.
The uncertainties in the table are statistical and energy-uncorrelated systematic uncertainties added in quadrature.
The day-night asymmetry (see Eq.\ (\ref{adn_eq}) ) of the recoil electron ES rate measured by SK is provided in Table \ref{adn_table_1} for each experimental phase, along with the analysis threshold and the day and night livetime.
This data set contains $\sim$4500 days of solar neutrino data, and it is the same data set analyzed to produce the solar neutrino oscillation analysis results reported in \cite{SK4Solar}.
However, the inclusion of the SK solar neutrino day-night asymmetry measurement is different between the SK solar neutrino oscillation analysis and the analysis presented in this work. 
Details about this difference will be discussed in Sec. \ref{sub:analysis_adn}.

In addition to using SK data to constrain effective NSI parameters, the SK measurement will be combined with both SNO and Borexino measurements.
A simultaneous fit between SK and SNO using the polynomial fit to the results for solar neutrino $P_{ee}(E_{\nu})$ and $A_\textrm{DN}(E_{\nu})$ from the 3 phases of SNO \cite{snob8paper,*snoheppaper,*sno3phase} is first performed, and subsequently combined with a fit to the Borexino $A_\textrm{DN}^\textrm{Be7}$ measurement.
The SK+SNO + Borexino $A_\textrm{DN}^\textrm{Be7}$ fit is denoted as ``Combined'' in the results section of this work.
\begin{table*}[]
\caption{The solar neutrino recoil electron spectra from the SK-I, SK-II, SK-III, and SK-IV 1664-day solar analyses in the ratio of the extracted number of events to the predicted unoscillated number of events. The uncertainty is statistical plus energy-uncorrelated systematic errors added in quadrature.}
\label{recoil_e_spec_table}
\center
\begin{tabular*}{\textwidth}{l @{\extracolsep{\fill}} c c c c }
\hline
\hline
$T_e$ Bin [MeV] & SK-I & SK-II & SK-III & SK-IV (1664 days) \\
\hline
 3.49-3.99 & $-$ & $-$ & $-$ & $0.4596^{+0.0596}_{-0.0582}$\\
 3.99-4.49 & $-$ & $-$ & $0.4476^{+0.1002}_{-0.0962}$ & $0.4151^{+0.0304}_{-0.0297}$  \\
 4.49-4.99 & $0.4529^{+0.0430}_{-0.0416}$ & $-$ & $0.4715^{+0.0579}_{-0.0557}$ & $0.4894^{+0.0235}_{-0.0230}$  \\
 4.99-5.49 & $0.4297^{+0.0229}_{-0.0225}$ & $-$ & $0.4200^{+0.0388}_{-0.0373}$ & $0.4515^{+0.0145}_{-0.0143}$  \\
 5.49-5.99 & $0.4491^{+0.0184}_{-0.0180}$ & $-$ & $0.4569^{+0.0350}_{-0.0335}$ & $0.4307^{+0.0120}_{-0.0118}$  \\
 5.99-6.49 & $0.4436^{+0.0154}_{-0.0151}$ & $-$ & $0.4327^{+0.0231}_{-0.0224}$ & $0.4423^{+0.0147}_{-0.0146}$  \\
 6.49-6.99 & $0.4606^{+0.0157}_{-0.0154}$ & $0.4386^{+0.0498}_{-0.0477}$ & $0.5037^{+0.0248}_{-0.0241}$ & $0.4456^{+0.0151}_{-0.0149}$  \\
 6.99-7.49 & $0.4758^{+0.0163}_{-0.0160}$ & $0.4476^{+0.0430}_{-0.0411}$ & $0.4244^{+0.0237}_{-0.0229}$ & $0.4394^{+0.0154}_{-0.0152}$  \\
 7.49-7.99 & $0.4567^{+0.0166}_{-0.0162}$ & $0.4609^{+0.0372}_{-0.0356}$ & $0.4673^{+0.0243}_{-0.0234}$ & $0.4544^{+0.0140}_{-0.0137}$  \\
 7.99-8.49 & $0.4306^{+0.0168}_{-0.0164}$ & $0.4729^{+0.0364}_{-0.0348}$ & $0.4686^{+0.0258}_{-0.0247}$ & $0.4387^{+0.0145}_{-0.0142}$  \\
 8.49-8.99 & $0.4536^{+0.0180}_{-0.0175}$ & $0.4633^{+0.0356}_{-0.0340}$ & $0.4200^{+0.0264}_{-0.0250}$ & $0.4433^{+0.0154}_{-0.0150}$  \\
 8.99-9.49 & $0.4635^{+0.0193}_{-0.0187}$ & $0.4987^{+0.0384}_{-0.0365}$ & $0.4443^{+0.0289}_{-0.0274}$ & $0.4306^{+0.0164}_{-0.0158}$  \\
 9.49-9.99 & $0.4561^{+0.0206}_{-0.0199}$ & $0.4742^{+0.0385}_{-0.0364}$ & $0.4234^{+0.0307}_{-0.0287}$ & $0.4248^{+0.0178}_{-0.0171}$  \\
 9.99-10.49 & $0.4087^{+0.0214}_{-0.0206}$ & $0.4811^{+0.0414}_{-0.0392}$ & $0.5294^{+0.0374}_{-0.0349}$ & $0.4060^{+0.0192}_{-0.0183}$  \\
 10.49-10.99 & $0.4718^{+0.0255}_{-0.0243}$ & $0.4525^{+0.0425}_{-0.0399}$ & $0.4810^{+0.0407}_{-0.0373}$ & $0.4305^{+0.0224}_{-0.0214}$  \\
 10.99-11.49 & $0.4387^{+0.0280}_{-0.0264}$ & $0.4693^{+0.0457}_{-0.0426}$ & $0.3912^{+0.0444}_{-0.0398}$ & $0.4585^{+0.0260}_{-0.0246}$  \\
 11.49-11.99 & $0.4603^{+0.0330}_{-0.0307}$ & $0.4824^{+0.0521}_{-0.0483}$ & $0.4785^{+0.0553}_{-0.0493}$ & $0.4207^{+0.0289}_{-0.0270}$  \\
 11.99-12.49 & $0.4654^{+0.0393}_{-0.0361}$ & $0.4194^{+0.0537}_{-0.0489}$ & $0.4245^{+0.0611}_{-0.0527}$ & $0.4223^{+0.0346}_{-0.0318}$  \\
 12.49-12.99 & $0.4606^{+0.0477}_{-0.0431}$ & $0.4617^{+0.0632}_{-0.0567}$ & $0.4003^{+0.0733}_{-0.0611}$ & $0.4454^{+0.0428}_{-0.0388}$  \\
 12.99-13.49 & $0.5819^{+0.0640}_{-0.0571}$ & $0.4437^{+0.0703}_{-0.0616}$ & $0.4223^{+0.0930}_{-0.0744}$ & $0.4618^{+0.0547}_{-0.0485}$  \\
 13.49-14.49 & $0.4747^{+0.0593}_{-0.0524}$ & $0.4303^{+0.0658}_{-0.0586}$ & $0.6630^{+0.1101}_{-0.0926}$ & $0.4857^{+0.0541}_{-0.0481}$  \\
 14.49-15.49 & $0.7236^{+0.1203}_{-0.1015}$ & $0.5625^{+0.1003}_{-0.0867}$ & $0.7134^{+0.2007}_{-0.1503}$ & $0.4113^{+0.0891}_{-0.0727}$  \\
 15.49-19.49 & $0.5747^{+0.1731}_{-0.1304}$ & $0.6477^{+0.1226}_{-0.1035}$ & $0.2124^{+0.2477}_{-0.1215}$ & $0.3447^{+0.1416}_{-0.1010}$  \\
\hline
\hline
\end{tabular*}
\end{table*}

\begin{table}[h!]
\caption{The lower energy threshold (in MeV) for the day-night asymmetry analysis in each SK-Phase and the resulting measurement for SK phase from the separate measurements of the day and night rates, including the statistical and systematic error (third column) and the time of exposure (in days) for the day (D) and night (N) periods (last column). The maximum bound on the recoil electron kinetic energy in the day-night asymmetry analysis  is 19.5 MeV for each experimental phase.}
\begin{tabular}{ c c c c }
\hline
\hline
Phase & $E_\textrm{thresh}$ & $A_\textrm{DN} \pm (\textrm{stat})\pm (\textrm{syst})$ & D/N exposure [days]\\
\hline
SK-I & 4.49 & $(-2.1\pm2.0\pm1.3)\%$ & 733.2/763.0 \\ 
SK-II & 6.49 & $(-5.5\pm4.2\pm3.7)\%$&383.6/408.2 \\
SK-III & 4.49 & $(-5.9\pm3.2\pm1.3)\%$ &264.1/284.5\\
SK-IV & 4.49 & $(-4.9\pm1.8\pm1.4)\%$ &797.4/866.5\\ 
\hline
\hline
\end{tabular}
\label{adn_table_1}
\end{table}

\subsection{
  \label{sub:nuprop} 
 Neutrino Propagation in Matter
}
When neutrinos propagate through matter, they experience a potential due to the electrons within. 
This matter potential takes the form 
\begin{equation}
\label{nu_mat_potential}
H_{mat} = \sqrt{2}G_{F}n_{e} 
\begin{bmatrix}
1 & 0 & 0 \\
0 & 0 & 0 \\
0 & 0 & 0
\end{bmatrix},
\end{equation}
where $n_{e}$ is the local electron number density and $G_{F}$ is the Fermi constant, and the rows and columns of $H_{mat}$ are both indexed as $e$, $\mu$, and $\tau$. 
All neutrinos produced in the Sun are created in the electron flavor state, and the probability of a neutrino exiting the Sun in a specific mass eigenstate is determined by propagating the neutrino through the Sun while determining the amplitude of the two flavor  
states at each step in the propagation.
During this propagation, one must take into account the matter potential's effect on the neutrino wavefunction due to the local electron density, and, in the case of NSI, the local quark densities.

For solar neutrinos, one can approximate their propagation in the regime where only one of the neutrino masses dominates, i.e. taking
$\Delta m_{31}^{2} \rightarrow \infty$, leaving a single relevant squared mass splitting $\Delta m_{21}^{2}$ \cite{two_flavor_approx_arg}. 
After rotating the Hamiltonian by the atmospheric mixing angle $\theta_{23}$ in the $\mu-\tau$ subspace and taking the upper left $2 \times 2$ sub matrix (see \cite{Friedland04} and \cite{3flavNSI2013}), the standard form of the two-neutrino Hamiltonian is obtained: 
\begin{equation}
\label{nu_2f_mat_potential}
H^{2f}_{mat} = \sqrt{2}G_{F}n_{e} 
\begin{bmatrix}
\cos^{2}\theta_{13} & 0 \\
0 & 0
\end{bmatrix},
\end{equation}
The validity of the two neutrino flavor approach is satisfied if $\theta_{13}\ll1$. 
In the analysis presented here, we fix the analysis bin of $\theta_{13}$ such that $\sin(\theta_{13}) = 0.02$. 
This $\theta_{13}$ bin is the closest to the best fit results from Daya Bay  \cite{daya_bay_2017,*daya_bay_hydrogen,*dayabay}, 
 RENO \cite{reno_2017, *reno}, and Double Chooz \cite{doublechooz_2014, *doublechooz}.

The local electron density $n_{e}$ is determined from the electron density profile of the Sun, given as a function of the solar radius by SSMs.
The SSMs also contain radial profile distributions of the solar density, the radiochemical mass fractions, and fractions of the total neutrino flux for each solar neutrino ``species.'' 
Here, the term species is used to distinguish between solar neutrinos due to the parent fusion processes that produce said neutrinos, such as the characteristic differences in the $P_{ee}$ between $hep$ and $^{8}$B solar neutrino species.
The radial profiles of the radiochemical mass fractions, along with the mass density profile and the assumption of the Sun as electrically neutral, are used to determine the actual electron and u and d quark number densities directly. 
In this analysis, the BP2004 SSM \cite{bp04} for calculating the solar neutrino probabilities is assumed. 
The $^{8}$B neutrino spectrum published by Winter et al.\ \cite{winter06} and the $hep$ solar neutrino spectrum from Bahcall et al.\ \cite{bahcall_hep_cx} are used to create the SK recoil electron spectral predictions.
These are the standard set of assumptions for the SK solar analyses.
The impact of the choosing the BP2004 SSM for this analysis as opposed to other SSMs was checked using the BP2000 SSM \cite{bp2000}, the BS2005 SSM \cite{bs05}, and the new B2016 SSMs given in \cite{new_ssms}. 
The solar model choice was found to be negligible in this analysis because the NC measurement of $^{8}$B neutrinos by SNO \cite{sno3phase} is used to constrain the solar neutrino flux through the nuisance parameter $\beta$ during the spectral fit described in Sec. \ref{sub:spec_pred}.

After the neutrino exits the Sun, each mass eigenstate is assumed to propagate incoherently  
as they travel to a detector on Earth.
For experiments that detect neutrinos via ($\nu$-$e$) ES, such as in the water in SK, 
D$_{2}$O in SNO, and the liquid scintillator in Borexino, the $\nu_{e}$ survival probability $P_{ee}$ must be determined.
If the neutrino arrives during the day time, i.e. the Sun is above the horizon, then the effective 2-flavor $P^{2f}_{ee}$ is directly obtained by rotating directly into the electron-basis with the solar mixing angle: 
\begin{equation}
P^{2f\textrm{ DAY}}_{ee} = \cos^{2}\theta_{12} - \cos(2\theta_{12})P_{2}(\cos^{2}\theta_{13}),
\end{equation} 
where $P_{2}(\cos^{2}\theta_{13})$ is the probability the electron neutrino produced in the solar core exits the Sun in the second mass state $\nu_{2}$. 
This calculation includes the effects from $\theta_{13}$ on the neutrino propagation through the matter potential in Eq.\ (\ref{nu_2f_mat_potential}).
Finally, one follows the standard procedure to convert to the effective two-flavor survival probability  $P_{ee}^{2f}$ to the three-flavor survival probability $P_{ee}$ (equation 17 from Fogli et al.\ (2000) \cite{p2_to_p3_nuosc_ref}): 
\begin{equation}
\label{Pee_2to3flavor}
P_{ee} =  \sin^{4}\theta_{13} + \cos^{4}\theta_{13} P_{ee}^{2f},
\end{equation}
where, again, $P_{ee}^{2f}$ retains the $\theta_{13}$ correction to the electron density in Eq.\ (\ref{nu_2f_mat_potential}): $n_{e}\cos^{2}\theta_{13}$.

If the neutrino traverses through the Earth during the night on its way to a detector, the Earth matter can effect the probability, including causing a regeneration of $\nu_{\mu/\tau}$ neutrinos into the $\nu_{e}$ state, as is the case with the LMA region in the case of no NSI.
If the incident neutrino wavefunction starts in the $\nu_{2}$ mass state and interacts with electrons in a detector 
in the $\nu_{e}$ flavor state, the transition probability $P_{2e}$ must be determined by tracing the neutrino wavefunction through the matter density profile of the Earth along a given zenith direction $z$. 
The definition of the zenith direction will be discussed later in section \ref{sub:pee_calc}.
In order to model the terrestrial matter potential properly, the Preliminary Reference Earth Model (PREM) \cite{prem_model} is used under the assumption of spherically symmetric Earth. 
The PREM table contains the radial profile of the Earth's electron density, and the symmetry approximation allows for the $P_{2e}^{2f,z}$ tables to be calculated only once for both the SK and SNO predictions.
\begin{equation}
\label{pee_zen}
P_{ee}^{2f,z} =  P_{1}P_{1e} + P_{2}P_{2e} = (1-P_{2})(1-P_{2e}^{z}) + P_{2}P_{2e}^{z}
\end{equation}
Once calculated, the $P_{2e}^{z}$ is combined with $P_{2}$, Eq.\ (\ref{pee_zen}), 
to obtain the $P_{ee}^{2f,z}$ for each zenith bin in the two-flavor approximation, 
then converted to the three-flavor $P_{ee}$ via Eq.\ (\ref{Pee_2to3flavor}).

\subsection{\label{sec:nsi}Non-Standard Interactions}
When the matter potential in the Hamiltonian is expanded to include NSI, it becomes
\begin{equation}
\label{nsi_hamiltonian}
H_{mat}^{NSI} = \sqrt{2}G_{F}n_{e} 
\begin{bmatrix}
1 + \epsilon_{ee} & \epsilon_{e\mu} & \epsilon_{e\tau} \\
\epsilon_{\mu e}* & \epsilon_{\mu\mu} & \epsilon_{\mu\tau} \\
\epsilon_{\tau e}* & \epsilon_{\tau\mu}* & \epsilon_{\tau\tau}
\end{bmatrix},
\end{equation}
where the diagonal (off-diagonal) terms of $H_{mat}^{NSI}$ are real (complex) \cite{Friedland04}\cite{3flavNSI2013}. 
This parameterization of the matter potential is described by an effective four-fermion operator
\begin{equation}
\label{nsi_lagrangian}
\mathcal{L}_{NSI} = -2\sqrt{2}G_{F} \epsilon_{\alpha\beta}^{f\textrm{ }P}(\bar{\nu}_{\alpha}\gamma^{\mu}\nu_{\beta})(\bar{f}\gamma_{\mu}Pf),
\end{equation}
where $P$ is the parity operator $(P = L,R)$, $f$ corresponds to fermions within the traversed matter ($e$, $u$, or $d$), and $\alpha,\beta$  = $e,\mu,\tau$. 
Because only the vector couplings ($V$) 
affect the neutrino propagation, the $L$ and $R$ components of the NSI parameters 
can be combined: $\epsilon^{f}_{\alpha\beta} = \epsilon^{fV}_{\alpha\beta} = \epsilon^{fR}_{\alpha\beta} + \epsilon^{fL}_{\alpha\beta}$.
Here, the $V$ term will be suppressed for simplicity.
The overall strength due to the NSI $\epsilon_{\alpha\beta}$ parameters in $H_{mat}^{NSI}$ is given by the sum of the contributions from each of the fermions in the matter, scaled by the ratio of the local fermion number density $n_{f}$ to $n_{e}$  respectively:
\begin{equation}
\label{comb_epsilon_eq}
\epsilon_{\alpha\beta} = \epsilon_{\alpha\beta}^{e} + \epsilon_{\alpha\beta}^{d} Y_{d} + \epsilon_{\alpha\beta}^{u}Y_{u} = \sum_{f = e,u,d} Y_{f} \epsilon^{f}_{\alpha\beta},
\end{equation}
where
\begin{equation}
\label{Yratios}
Y_{f} = \frac{n_{f} } {n_{e}} \quad (f = e, u, d).
\end{equation} 
Since the different fermion epsilons are allowed to take on differing values relative to one another, this complicates matters greatly due to the large parameter space. 

Because we can approximate solar neutrino oscillations in a two-flavor scenario, and to simplify the parameter space by including NSI, we reduce the NSI matter potential from Eq.\ (\ref{nsi_hamiltonian}) to the two-flavor approximation.
Then, the effective NSI matter potential is 
\begin{equation}
\label{nu_nsi_matpot}
H^{\textrm{NSI}}_{mat} = \frac{G_{F}n_{e}}{\sqrt{2}}
\begin{bmatrix}
\cos^{2}\theta_{13} + \epsilon_{11} & \epsilon_{12} \\
\epsilon_{12}* & -\cos^{2}\theta_{13} - \epsilon_{11}
\end{bmatrix},
\end{equation}
where we follow the notation convention for the two-flavor NSI parameters of $\epsilon_{11}$ and $\epsilon_{12}$ established by Friedland et al.\ in \cite{Friedland04}.
However, we use the three neutrino derivation from the work of Gonzalez-Garcia and Maltoni \cite{3flavNSI2013} as well as their choice of the conjugation of $\epsilon_{12}^{f}$.
In this notation, the total possible contributions from each fermion to the total strength of the NSI parameter $\epsilon_{1j}$ is
\begin{equation}
\label{comb_epsilon_eq_mb} 
\epsilon_{1j} = \epsilon_{1j}^{e} + \epsilon_{1j}^{d} Y_{d} + \epsilon_{1j}^{u}Y_{u} = \sum_{f = e,u,d} Y_{f} \epsilon^{f}_{1j},
\end{equation}
where $j = 1,2$, and $Y_{f}$ is the same electron-to-fermion ratio defined in Eq.\ (\ref{Yratios}). 

Following the work of Gonzolez-Garcia and Maltoni \cite{3flavNSI2013}, we can explicitly write the effective NSI parameters $\epsilon_{1j}^f$ in terms of the PMNS parameters and $\epsilon_{\alpha\beta}^f$ NSI parameters:
\begin{equation}
\label{ggm_eD}
\begin{matrix}
-\frac{1}{2}\epsilon_{11}^f = c_{13}s_{13}\textrm{Re}[e^{i\delta_{\textrm{CP}}}(s_{23}\epsilon_{e\mu}^{f}+ c_{23}\epsilon_{e\tau}^{f})] \\  \vspace{0.05cm} \\
- (1+s^{2}_{13})c_{23}s_{23}\textrm{Re}[\epsilon_{\mu\tau}^{f}]\\  \vspace{0.05cm} \\
- \frac{c_{13}^{2}}{2}(\epsilon_{ee}^{f} - \epsilon_{\mu\mu}^{f})
+ \frac{s^{2}_{23} - s^{2}_{13}c^{2}_{23}}{2}(\epsilon_{\tau\tau}^{f} - \epsilon_{\mu\mu}^{f})
\end{matrix}
\end{equation}
and
\begin{equation}
\label{ggm_eN}
\begin{matrix}
\frac{1}{2}\epsilon_{12}^f  =  c_{13}(c_{23}\epsilon_{e\mu}^{f} - s_{23}\epsilon_{e\tau}^{f}) \\ \vspace{0.05cm} \\
+ s_{13} e^{-i\delta_{\textrm{CP}}}[s_{23}^{2}\epsilon_{\mu\tau}^{f} - c_{23}^{2}\epsilon_{\mu\tau}^{f*} + c_{23}s_{23}(\epsilon_{\tau\tau} - \epsilon_{\mu\mu}^{f})],
\end{matrix}
\end{equation}
where $s_{ij}$ and $c_{ij}$ are $\sin\theta_{ij}$ and $\cos\theta_{ij}$ respectively.
Taking muon-neutrino-related terms and $\theta_{13}$ to zero, the equations for $\epsilon^{f}_{11}$ and $\epsilon^{f}_{12}$ given in Friedland et al.\ \cite{Friedland04} are recovered.
This reduction of the parameter space to only two additional parameters from NSI allows for a reasonable way to test NSI experimentally.
While the condition that $\theta_{13} \ll 1$ for the two-flavor approximation is still satisfied, there is an additional condition that, in the presence of NSI, $G_{F}\Sigma_{f}\epsilon_{\alpha\beta}^{f} \ll \Delta m_{31}^{2}/E_{\nu}$.  
In the approximation of the $\nu_{3}$ single mass domination $\Delta m_{31}^{2} \rightarrow \infty$, this additional condition is satisfied.

As described in \cite{Friedland04}, CC reactions, such as those that produce solar neutrinos, remain unchanged by Eq.\ (\ref{nsi_lagrangian}).
However, ES reactions can be affected by the extra coupling with electrons ($\epsilon_{\alpha\beta}^{e}$ or $\epsilon_{1j}^{e}$), which would modify the cross-sections through both the axial and vector currents. 
This means that the CC reactions that detectors like the radiochemical experiments, SNO, and KamLAND see will remain unchanged by the introduction of NSI. 
Conversely, the neutral current (NC) reaction in SNO could be affected by NSI as it is dependent on the axial current, and the ES neutrino-electron cross-sections and detection reactions in SK, Borexino, and SNO (D$_{2}$O) would be affected by $\epsilon_{\alpha\beta}^{e}$.

In this analysis, any extra contribution due to NSI with electrons in matter is neglected $(\epsilon_{1j}^{e} = 0)$, so that the ES cross sections remain unaffected.
 Only the NSI between solar neutrinos and a single quark ($u$ or $d$) is considered at a time: 
i.e. either $\epsilon_{1j} = \epsilon_{1j}^{d}Y_{d}$ in the case of ($\nu$-$d$) NSI, or $\epsilon_{1j} = \epsilon_{1j}^{u}Y_{u}$ in the case of ($\nu$-$u$) NSI. 
We present our results in terms of the effective NSI parameters for $\epsilon_{1j}^{f}$ for $f=u, d$ and $j =$ 1, 2.
 As this analysis uses the NC constraint from SNO as well as their data, any extracted dependence of the $\epsilon_{\alpha\beta}^{f}$ ($f = u, d$) from the effective NSI parameters $\epsilon_{1j}^f$ would need to be only the vector component $\epsilon_{\alpha\beta}^{V,f}$. 
 The axial component would need to be set to zero $\epsilon_{\alpha\beta}^{A,f}=0$.
 This procedure is the same as the method described in \cite{3flavNSI2013}.
%

The effective mass squared splitting $\Delta m_{\textrm{eff}}^{2}$ and effective mixing angle $\theta_{\textrm{eff}}$ in the presence of matter including NSI can be determined by diagonalizing the NSI matter potential given in Eq.\ (\ref{nu_nsi_matpot}). These effective oscillation parameters take the form
\begin{equation}
\label{eff_nsi_dm2_theta}
\Delta m_{\textrm{eff}}^{2} = \sqrt{ K^{2} + \Sigma^{2} }\quad
\textrm{and}\quad \tan(2\theta_{\textrm{eff}}) = \frac{\Sigma}{K}
\end{equation}
where
\begin{equation}
K = A(\cos^{2}\theta_{13} + \epsilon_{11}^{f}Y_f) - \Delta m_{21}^{2}\cos(2\theta_{12}),
\end{equation}
\begin{equation}
\Sigma = A\epsilon_{12}^{f}Y_f + \Delta m_{21}^{2}\sin(2\theta_{12}),
\end{equation}
and $A = 2\sqrt{2}G_{F}n_{e}E_{\nu}$.
This reduces to the SI matter mixing angle and matter mass squared splitting by setting $\epsilon_{11}^f=\epsilon_{12}^f=0$.

Current experimental constraints to flavor-dependent NSI $(\epsilon_{\alpha\beta}^{f})$ have been performed by SK \cite{sk_atm_nsi} and by IceCube DeepCore \cite{icecube_deepcore_nsi_2018} using atmospheric neutrino data, as well as by the neutrino beam experiments NuTeV \cite{nutev, *nutev_er} and CHARM-II \cite{charm}.
These results constrain the $\mu$-related terms to be close to zero, i.e. $\epsilon_{\mu\alpha} = \epsilon_{\beta\mu} \simeq 0$ \cite{Friedland04} \cite{icecube_deepcore_nsi_2018}.
Additional constraints have been published by the COHERENT neutrino-nucleus elastic scattering experiment in \cite{coherent} for electron-flavor vector-coupled NSI parameters: ($\epsilon_{ee}^{dV}$, $\epsilon_{ee}^{uV}$). 
These constraints can be used in conjunction with the results for ($\epsilon_{11}^{f}$,
$\epsilon_{12}^{f}$) from the work presented in this paper to obtain limits on the NSI parameters $\epsilon_{\alpha\beta}^{f}$ for $f = u, d$.
This is beyond the results of this analysis, which focuses solely on constraints solar neutrino data place on $\epsilon_{11}^{f}$ and $\epsilon_{12}^{f}$ for $f = u$ or $d$.

With these choices, the effect that the effective NSI parameters have on the recoil electron spectrum seen by SK, SNO's measured neutrino spectrum, and Borexino's $A_\textrm{DN}$ measurement of $^{7}$Be solar neutrinos, are determined through calculating the probabilities $P^{2f}_{2}, P_{2e}^{2f,z}$, and $P_{ee}$ in the same method described for the SI case in the beginning of this section.
\section{\label{sec:analysis}Analysis Details}
\subsection{\label{sub:pee_calc}Probability calculations}
Although an analytical approximation is used in \cite{Friedland04}, the approximation breaks down for the NSI equivalent of the small angle MSW effect when adiabaticity of neutrino flavor conversion is violated.
This situation occurs for negative values of $\epsilon_{12}$ and $|\theta_{12} - \frac{1}{2}\tan^{-1}(|\epsilon_{12}|/[1+\epsilon_{11}])| \ll 1$.
In order to accurately determine the effect NSI has on the solar neutrino $P_{ee}$'s for the full range of oscillation and NSI parameters used in this analysis,  the neutrino wavefunction was traced by stepping through the Sun and the Earth.
The neutrino wavefunction and the subsequent oscillation probabilities are dependent on the set of oscillation parameters $(\epsilon_{11}^{f},\epsilon_{12}^{f},\theta_{13},\theta_{12},E_{\nu}/\Delta m_{21}^{2})$.

\begin{figure}[h]
\includegraphics[scale=.33]{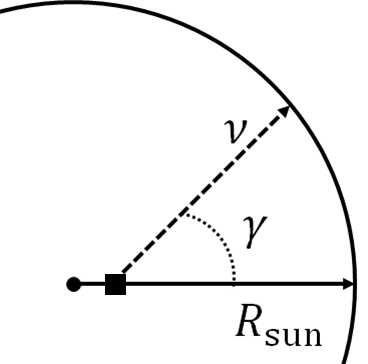}
\caption{Illustration of the angle $\gamma$ between the solar neutrino direction (dashed line) and the solar radius vector (solid ray), with the neutrino creation point denoted by the square.}
\label{fig:solar_angle_gamma_deff} 
\end{figure}

When tracing the neutrino wavefunction through the Sun, the solar radius $R_{\textrm{Sun}}$ is used to form production bins $0.02\times R_{\textrm{Sun}}$ wide.
These radial bins span from the core to $0.32\times R_{\textrm{Sun}}$, with the neutrino produced at and propagated from the center of each radial bin.
The fractional contribution of the total neutrino production 
for a given solar neutrino species $s$ is determined for each of these radial bins, indexed by $i$, and a weight $f_{i}^{s}$ is assigned based on these fractions. 
In each of these radial bins, the neutrino wavefunction is produced in the $\nu_{e}$ state, and then propagated along 11 directions corresponding to $\cos\gamma$ = $[-1.0, -0.8, ..., 0.8, 1.0]$, the dot product between the neutrino direction and the radial vector originating at the Sun's core.
The definition of the angle $\gamma$ is illustrated in Fig. \ref{fig:solar_angle_gamma_deff}.
For $^{8}$B neutrinos, which are typically produced around $0.04\times R_{\textrm{Sun}}$, the minimum (maximum) pathlength through the solar core, where 
$R_{\textrm{core}} \simeq 0.19 R_{\textrm{Sun}}$, is approximately $1.04\times10^{5}$ $(1.60\times10^{5})$ km, or 8.19 (12.6) times the diameter of the Earth.

After propagating the neutrino wavefunction through the Sun, the probability of a solar neutrino species exiting the Sun for a given radial bin $i$ in, say, the second mass eigenstate $P_{2,i}$, is determined by averaging the probability over the 11 $\cos\gamma$ directions.
The average $P_{2}^{s}$ for each solar neutrino species $s$ ($^{8}$B, $hep$, and $^{7}$Be neutrinos) is then determined by averaging the $P_{2,i}$ with the $f_{i}^{s}$ weights:
\begin{equation}
P^{s}_{2}(\epsilon_{11}^{f},\epsilon_{12}^{f},\theta_{13},\theta_{12},E_{\nu}/\Delta m_{21}^{2}) = 
\sum_{i} f_{i}^{s} P_{2,i}.
\end{equation}
Once entering the vacuum of space outside of the Sun, the neutrino mass eigenstate is propagated incoherently, reaching Earth in the same mass eigenstate with which it exits the Sun.

For tracing the neutrino wavefunction along a zenith direction through matter in the Earth, the neutrino wave function is started in the $\nu_{2}$ mass state in order to later combine it with the $P_{2}^{s}$ of each solar neutrino species. 
A single zenith bin is used for the day-time predictions for SK and SNO, while the night region is subdivided evenly into 1000 bins of $\cos\theta_{z}$.  
SK, SNO and Borexino have accurate timing for their events, and SK and SNO are able to reconstruct the incident neutrino direction.  
Using this information combined with the fact that the position of the Sun is known incredibly accurately means that the zenith dependence for the exposure of each detector is important when predicting the rate that should appear in a detector relative to a zenith direction, or the cumulative rate seen by a detector during the night.
The zenith direction is defined by $\cos\theta_{z}$, where the angle $\theta_{z}$ is defined by the dot product of the upward vertical direction of a detector with the incident neutrino direction.
With this definition, the zenith angle $\theta_{z}$ is actually the Nadir angle, as $\cos\theta_{z}$ is negative for incident solar neutrinos during the day and $\cos\theta_{z}$ is positive for solar neutrinos coming up from the Earth during the night, with the extreemum of $\cos\theta_{z}$ occurring along the radial vector of the Earth's spherical coordinate system. 
\begin{table}[h!]
\caption{Relevant information from the Preliminary Reference Earth Model \cite{prem_model} for determining the matter potential of neutrinos propagating through the Earth's mantle and/or core. The ratio $Y_{e}$ of electrons to the matter density $\rho$ is given in the second column. The third (fourth) column contains the ratio $Y_{u} (Y_{d}) = n_{u (d)} / n_{e}$ of the up (down) quark number density to the electron number density.}
\begin{tabular}{ c c c c }
\hline
\hline
 Region & $Y_{e}$ [mol/g] & $Y_{u}$ & $Y_{d}$\\
\hline
Mantle & 0.497 & 3.012 & 3.024\\
Core & 0.468 & 3.137 & 3.274\\
\hline
\hline
\end{tabular}
\label{prem_yfactor_table}
\end{table}

In order to propagate the neutrino wavefunction through the Earth along a given zenith direction, regions of constant density are determined beforehand via interpolation of the Preliminary Reference Earth Model (PREM) table \cite{prem_model}. 
These regions are defined so that if the density fluctuates more than a certain amount relative to the current density, 0.001 g/cm$^{3}$, the current density region is ended with its length recorded, and a new density region is used.
Because the electron and quark densities are different than their values in the mantle region, the points at which the neutrino trajectory enters and leaves the Earth's core is tracked, though this only occurs for $\cos\theta_{z}\ge 0.883$. 
From the PREM model, the ratio of electrons to the matter density in the Earth is determined through assumptions of the Earth's chemical composition in the mantle and core. 
For the mantle (core) this ratio is $Y_{e} =$ 0.497 (0.468) mol/g, which yields the electron number density after multiplying by the local density mass $\rho$ and Avogadro's number $N_{A} = 6.022\times10^{23}$. 
Additionally, the PREM model gives the ratio of neutrons to electrons as $Y_{n} =$ 1.012 (1.137) in the mantle (core). 
By assuming the Earth is electrically neutral, 
the values of $Y_{u}$ and $Y_{d}$ can be calculated for the mantle (core). 
These ratios are listed in Table \ref{prem_yfactor_table}. 
When propagating the neutrino wavefunction through the terrestrial matter, the local matter density in the mantle and core regions is multiplied by these factors in order to obtain the local fermion number densities that contribute to the matter potential.
\subsection{  \label{sub:spec_pred}SK Recoil Electron Spectrum Predictions and $\chi^{2}$ Fit}
This analysis uses the same $\chi^{2}$  fitting method to determine how well a set of solar neutrino oscillation parameters $(\sin^{2}\theta_{12},\Delta m_{21}^{2},\sin^{2}\theta_{13})$ will fit the SK-I/II/III/IV recoil electron spectrum as in previous SK solar neutrino oscillation results described in section VI of \cite{SK4Solar} and originally proposed in G.\ L.\ Fogli et al.\ (2002) \cite{fogli_chi2_for_solar}.
The inclusion of the SNO results \cite{sno3phase} into a simultaneous fit is also performed using the same methods as in the SI SK solar neutrino oscillation analysis.
The process of determining the spectral predictions for the ES recoil electron signature in SK is described in detail in \cite{SK1Solar}, but will be discussed here as well.
The spectral predictions are determined by calculating the unoscillated and oscillated event rates scaled by SK's exposure during the day and for each zenith bin Z during the night for both $^{8}$B and $hep$ solar neutrinos.
Though the dominant contribution to the signal in SK (and SNO) is due to $^{8}$B neutrinos, the $hep$ neutrinos have a small, non-negligible impact on the spectra at the higher energy bins.

The spectra are formed from the zenith-dependent $P^{\textrm{Z}}_{ee}$ by applying to them so-called "transfer matrices" \cite{minos_tfm} that incorporate the effects of the SK energy resolution, and the uncertainties from the neutrino spectrum and the ES cross sections. 
The transfer matricies propagate these effects onto the expected ES event rates 
due to solar neutrinos from both CC and NC processes.
The expected induced event rate of ES events in energy bin $e$ by $^{8}$B ($hep$) solar neutrinos is denoted by $B_{e,\textrm{Z},p}$ $(H_{e,\textrm{Z},p})$ in the case of no neutrino oscillations, for zenith bin Z and SK phase $p$.
In the case of oscillations, the expected ES event rate is denoted as $B_{e,\textrm{Z},p}^{\textrm{osc}}$ and $H_{e,\textrm{Z},p}^{\textrm{osc}}$. 
The expected recoil electron spectrum is formed by dividing the oscillated rate for each solar neutrino species by the total unoscillated rate due to both $^{8}$B and $hep$ solar neutrinos:
\begin{equation}
b_{e,\textrm{Z}}^{p} = \frac{B_{e,\textrm{Z},p}^{\textrm{osc}}}{B_{e,\textrm{Z},p}+H_{e,\textrm{Z},p}} \textrm{ and } h_{e,\textrm{Z}}^{p} = \frac{H_{e,\textrm{Z},p}^{\textrm{osc}}}{B_{e,\textrm{Z},p}+H_{e,\textrm{Z},p}}.
\end{equation}
Similarly, the measured rate $D_{e,\textrm{Z},p}$ is used to obtain the measured spectrum $d_{e,\textrm{Z}}^{p} = D_{e,\textrm{Z},p}/(B_{e,\textrm{Z},p}+H_{e,\textrm{Z},p})$ that the predictions are compared against, though here, the unoscillated $^8$B and $hep$ rates are determined by SK solar neutrino Monte Carlo simulation.
The day spectra are obtained from a unified daytime zenith bin, while the night spectra are formed by combining the rates of all night-time zenith bins for each energy bin.
Likewise, the total spectrum combines all zenith bins for each energy bin and SK phase separately to form $b_{e}^{p}$, $h_{e}^{p}$ for the predictions, and $d_{e}^{p}$ for the data.
Finally, the event rate predictions for the recoil electron spectra are $r_{e}^{p} = \beta b_{e}^{p} + \eta h_{e}^{p}$, where $\beta$ and $\eta$ are nuisance parameters that scale the  $^{8}$B and $hep$ rates respectively.
\begin{figure*}
\center
%
\includegraphics[width = 0.99\linewidth]{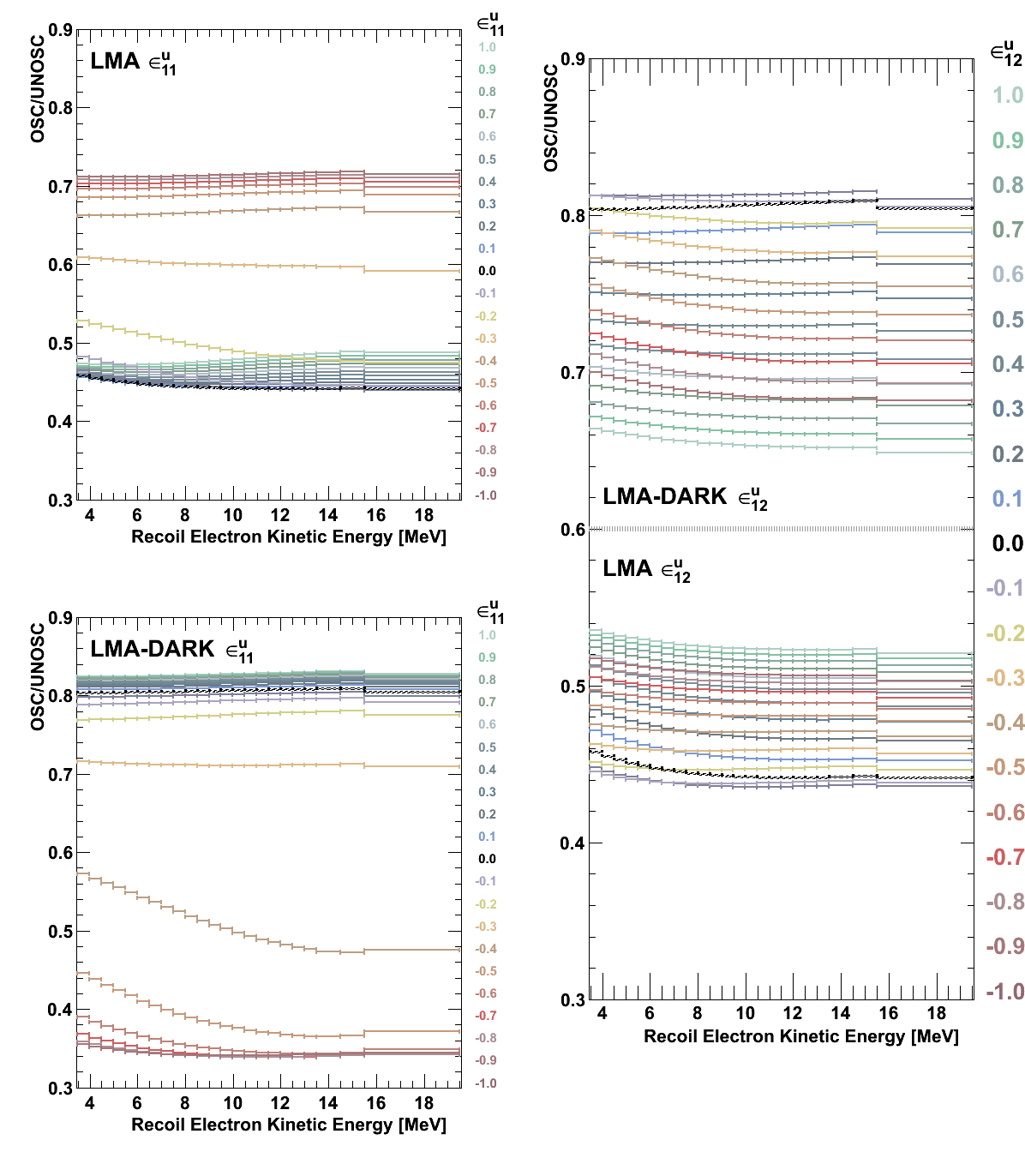}\\
\caption{Predictions of the SK-IV recoil electron spectrum due to $^8$B + $hep$ solar neutrinos for various NSI scenarios (color online). 
The left (right) column corresponds to varying $\epsilon_{11}^{u}$ ($\epsilon_{12}^u$) only.
The LMA region predictions correspond to the SI SK data LMA best-fit point ($\sin^2\theta_{12}$ =  0.334), and the LMA-DARK region corresponds to an equivalent darkside LMA-like point ($\sin^2\theta_{12}$ = 0.799). 
For each prediction, $\Delta m_{21}^{2} = 4.07 \times 10^{-5} \textrm{ eV}^2$ and all other NSI parameters are set to zero.
The SI spectrum (0.0) is plotted as a black dashed line in each case. 
In the right plot, the LMA and LMA-DARK predictions do not cross osc/unosc = 0.6 (gray dotted line).
}
\label{spectral_prediction_plots}
\end{figure*}

Various examples of SK-IV spectral predictions for ($\nu$-$u$) NSI are plotted in Fig. \ref{spectral_prediction_plots} for SI oscillation parameters from the LMA best-fit 
($\sin^2\theta_{12}$ = 0.0.334 and $\Delta m_{21}^{2} = 4.07 \times 10^{-5} \textrm{ eV}^2$) and an equivalent ``darkside''-LMA with $\sin^2\theta_{12}$ = 0.799.
The so-called ``darkside'' region refers to values of $\sin^2\theta_{12}$ greater than maximal mixing ($\sin^2\theta_{12} = 0.5$), while the so-called ``lightside'' region refers to values of $\sin^2\theta_{12}$ less than maximal mixing. 
The strength of the given NSI parameter is provided in color on the right side of each figure, and all other NSI parameters are set to zero.
The SI spectrum is plotted as a dashed black line.
Each spectra contains the contribution of both $^8$B and $hep$ neutrinos across all zenith bins.
As one can see from the right column, there is a sharp transition between $-0.4 < \epsilon_{11}^u < -0.2$ corresponding to a resonance region induced by the NSI parameter. 
This resonance causes behavior similar to maximal mixing $\sin^2 \theta_{12} = 0.5$ for SI, and causes the best-fit solution to transition from the LMA region in the lightside to a darkside LMA-like region for many NSI scenarios.
In the case of the NSI parameter $\epsilon_{12}^u$ (left side of Fig. \ref{spectral_prediction_plots}), the change in the spectrum is much smoother.
For the LMA (darkside-LMA) SI oscillation parameters, the larger magnitude values of $\epsilon_{12}^u$ shift the spectra up (down), though the sign affects the energy dependence (slope) and shape of the low-energy upturn.
While not shown here, the ($\nu$-$d$) NSI scenarios cause similar behaviors to the spectra, though the effect for each step in the value of $\epsilon_{1j}^d$ is slightly smaller.
This is due to the lower d-quark density compared to the u-quark density inside the Sun.

The $\chi^{2}_{\textrm{spec}}$ fits the predicted rates $r_{e}^{p}$ to the measurement $d_{e}^{p} \pm \sigma_{e}^{p}$ by applying the spectral distortion factor $f_{e}^{p}(\tau,\epsilon_{p},\rho_{p})$ derived from the energy-correlated systematic uncertainties (Figure 37 in \cite{SK4Solar}) to the calculated rate $r_{e}^{p}$. 
 A systematic shift in the energy scale or a deviation from the energy resolution in phase $p$ from the detector is described by the constrained nuisance parameters $\epsilon_{p}$ and $\rho_{p}$ respectively.
The uncertainty for the SK-IV absolute energy scale is $\pm$0.54\%, while the energy resolution uncertainty is $\pm$1.0\% below 4.89 MeV and
$\pm$0.6\% above 6.81 MeV \cite{SK4Solar}. 
Further details can be found in \cite{nakano_thesis}.
The constrained nuisance parameter $\tau$ describes a systematic shift of the $^{8}$B neutrino spectrum by $\pm$100 keV \cite{winter06}.
The systematic nuisance parameters $\tau,\rho_{p},$ and $\epsilon_{p}$ are standard Gaussian variables constrained to $0\pm1$.
When performing a combined SK fit that includes all experimental phases, the nuisance parameters $\beta, \eta$, and $\tau$ are applied simultaneously across each phase.

Given the dependence of the calculated rate in an energy bin of each phase on the NSI, oscillation, and nuisance parameters 
$r_{e}^{p} = r_{e}^{p}(\epsilon_{11}^{f},\epsilon_{12}^{f},\theta_{13},\theta_{12},\Delta m_{21}^{2}\textrm{; } \beta, \eta)$, 
the $\chi^{2}_{p}$ for a single phase that must be minimized is 
\begin{equation}
\chi_{p}^{2}(\beta,\eta) = \sum_{e=1}^{N_{p}} \frac{d_{e}^{p} - f^{p}_{e}r^{p}_{e}}{\sigma_{e}^{p}}
\end{equation}
and the minimization over the five nuisance parameters is
\begin{equation}
\label{chi2_min}\chi^{2}_{\textrm{spec,}p} 
= \min_{\tau,\rho_{p},\epsilon_{p},\beta,\eta}(\chi^{2} _{p,\textrm{data}} + \tau^{2} + \rho_{p}^{2} + \epsilon_{p}^{2} + \Phi).
\end{equation}
The flux constraints from prior uncertainties on $\beta,\eta$ are incorporated into the $\Phi$ term via $\Phi = (\beta - \beta_{0})^{2}/\sigma_{\beta}^{2} + (\eta - \eta_{0})^{2}/\sigma_{\eta}^{2}$. 
These priors are taken as to constrain $\beta$ to the $^{8}$B flux of $(5.25 \pm 0.20)\times10^{6}/$(cm$^{2}$ sec),  motivated by the SNO NC measurement of the total $^{8}$B flux \cite{sno3phase}, 
and a weak constraint on $\eta$ from a $hep$ flux of $(8\pm16)\times10^{3}/$(cm$^{2}$ sec) \cite{snoheppaper}.
In order to incorporate additional energy-independent systematic uncertainties on the total rate not covered by the spectral distortion factor $f_{e}^{p}$ or the statistical measurement error $\sigma_{e}^{p}$, an additional parameter $\alpha_{p}$ is introduced which scales the a posteriori constraints on the total rate parameters $\beta$ and $\eta$ by $\alpha_{p}^{-1/2}$ without affecting the minimum $\chi^{2}$.
The row labeled as Subtotal in Table V of \cite{SK4Solar} contains the total energy-independent uncertainties for $\alpha_{p}$, where each contributing uncertainty is integrated over all energies, and the parameter is chosen to have the form 
$\alpha_{p} = \sigma_{p,\textrm{stat}}^{2} / (\sigma_{p,\textrm{stat}}^{2} + \sigma_{p,\textrm{syst}}^{2})$.
Since Eq.\ (\ref{chi2_min}) is quadratic in $\beta$ and $\eta$, it can be expressed using the difference between the best-fit rate parameters $\beta_{\textrm{min}}^{p}$ ($\eta_{\textrm{min}}^{p}$) and $\beta$ ($\eta$):
\begin{equation}
\label{quad_form_chi2}
\chi^{2}_{\textrm{spec},\alpha_{p}} = 
\chi^{2}_{p,\textrm{min}} + 
\alpha_{p}\vec{\bold{\beta}}\cdot\bold{C}_{p}\cdot 
\vec{\bold{\beta}}^{\textrm{T}}
\end{equation}
where 
$\vec{\bold{\beta}} = (\beta - \beta_{\textrm{min}}^{p},\eta - \eta_{\textrm{min}}^{p})$,
 and $\bold{C}_{p}$ is a $2\times2$ curvature matrix used to find the minimum in the $(\beta,\eta)$ space.
Finally, the $\chi^{2}$ for the combined SK fit to the spectra and rates is
\begin{equation}
\label{tot_spec_x2}
\chi^{2}_{\textrm{spec}} = 
\min_{\tau,\rho_{p},\epsilon_{p},\beta,\eta}
\bigg( \tau^{2} +\Phi + \sum_{p=1}^{4}(\chi^{2} _{p,\alpha_{p}} + \rho_{p}^{2} + \epsilon_{p}^{2}) \bigg),
\end{equation}
where we use the quadratic form for $\chi^{2}_{\textrm{spec},\alpha_{p}}$ from Eq.\ (\ref{quad_form_chi2}).
The final rate nuisance parameters $\beta$ and $\eta$ from this combined fit will be used to predict the day-night asymmetries for each phase for $\chi^{2}$ fitting (Sec \ref{sub:analysis_adn}).

When considering the SK data alone, the SNO measurement of the $^{8}$B NC is used as a constraint on $\beta$.
This constraint is dropped when including the SNO data from \cite{sno3phase}. 
While both SK and SNO measure a similar energy range of $^{8}$B neutrinos, the $P_{ee}$ below $\sim$7 MeV is more tightly constrained by the SK recoil electron spectrum and day-night asymmetry measurement.
As SNO measures of the ratio of the CC to NC reactions between neutrinos and deuterium in D$_{2}$O, more higher energy events are detected due to the tighter energy correlation from the CC interaction.
Additionally, due to the energy dependence of the neutrino-deuterium NC and CC cross sections, SNO constrains $P_{ee}$ above $\sim$7 MeV more tightly than SK.
Thus both experiments complement the $^{8}$B solar neutrino $P_{ee}$ measurement from the other. 
In the SI case, SK constrains the $\Delta m_{21}^{2}$ parameter better than SNO due to a more precise measurement of the day-night asymmetry, while SNO has a tighter constraint on $\sin^{2}\theta_{12}$ due to their simultaneous measurement of the NC and CC rates.
The combined analysis benefits from correlations between the experiments, which leads to much better constraints to the oscillation (and NSI) parameters than can be achieved by simply adding the $\chi^{2}$ from each experiment together.

The combined analysis uses the same $hep$ flux constraint of $(7.9\pm1.2)\times10^{3}/$(cm$^{2}$ sec) as SNO \cite{sno3phase}, taken from the BS05 SSM \cite{bs05}.
The procedure used to analyze the SNO data in this analysis is the same as suggested by SNO and described in in \cite{sno3phase}.
They report the coefficients and correlation matrix of a quadratic fit of $P_{ee}^{\textrm{DAY}}(E_{\nu}-10 \textrm{ MeV})$ to their spectral data and of a linear fit to their $A_\textrm{DN}$ spectral measurement also centered around $E_{\nu} = 10$ MeV.
The results SNO provides for these fits are used when performing the combined the fit to SK data simultaneously with SNO (SK+SNO), which is the same analysis method used to perform the SK+SNO solar neutrino oscillation analysis for SI \cite{SK4Solar}.
The recoil electron spectrum data for each phase of SK is listed in Table \ref{recoil_e_spec_table}. 



\subsection{\label{sub:analysis_adn}Day-Night Asymmetries}
The day-night asymmetry $A_\textrm{DN}$ measurement by SK is treated differently here than in the standard SK solar oscillation analysis due to the computational demands required by increasing the overall parameter space when including the NSI parameters.
However, a simpler method of predicting $A_\textrm{DN}$ has been developed to include fits to the measured SK day-night rate asymmetries provided in Table \ref{adn_table_1}. 
In order to form the total rate during the day (D) and night (N) for use in this analysis, the total oscillated and unoscillated rates are obtained for a given zenith region Z (Z = D or N) by summing over the relevant energy range above the analysis threshold $E^{p}_\textrm{thresh}$ for a given phase $p$. 
Subsequently, one obtains
\begin{equation}
\label{rates_eq}
r_{\textrm{Z}}^{p} = \frac{\sum_{e=E_\textrm{thresh}^{p}}^{19.5 \textrm{MeV}}r^{p\textrm{,OSC}}_{e\textrm{,Z}}}{\sum_{e=E_\textrm{thresh}^{p}}^{19.5 \textrm{MeV}}r^{p\textrm{,UNOSC}}_{e\textrm{,Z}}},
\end{equation}
where the index $e$ corresponds to the energy bins of the recoil electrons in the analysis.
One can define $b^{p}_{\textrm{Z}} \equiv r_{\textrm{Z}}^{p}$ by only using the underlying $^{8}$B neutrinos rates for the predictions, and similarly defining $h^{p}_{\textrm{Z}}$ for $hep$ neutrino rates.

This procedure is done separately for the $^{8}$B and $hep$ day and night rates, in order to later combine them properly to form the day-night asymmetry predictions.
The lower energy threshold of the recoil electron kinetic energy for the day-night asymmetry analysis in SK-I, SK-II, and SK-IV is 4.49 MeV and 6.49 MeV in SK-II.
The details for each $A_\textrm{DN}^{p}$ measurement of SK-Phase $p$ is given in Table \ref{adn_table_1}, with the statistical and systematic errors given in that order, as well as the energy range and the time exposure of the day and night periods.

After obtaining the best fit flux and nuisance parameters $\beta$ and $\eta$ from the minimized $\chi^{2}$ fit to the combined SK data, they are combined with the total day and night rates for both $^{8}$B and $hep$ from Eq.\ (\ref{rates_eq}) in order to form the predictions of the day-night asymmetry (in fractional form) for each SK-Phase:
\begin{equation}
\label{int_adn}
A_\textrm{DN}^{p} = 2 \frac{ 
(b^{p}_{\textrm{D}} - b^{p}_{\textrm{N}})\beta + 
(h^{p}_{\textrm{D}} - h^{p}_{\textrm{N}})\eta 
}{
(b^{p}_{\textrm{D}} + b^{p}_{\textrm{N}})\beta + 
(h^{p}_{\textrm{D}} + h^{p}_{\textrm{N}})\eta }.
\end{equation}
Using these predicted $A^{p}_\textrm{DN}$, a simple $\chi^{2}$ fit to the measured straight asymmetries listed in Table \ref{adn_table_1}  is performed separately for each phase and is summed together with the combined SK fit to the spectrum and rate 
 $\chi_{\textrm{spec}}^{2}$ from Eq.\ (\ref{tot_spec_x2}):
\begin{equation}
\chi^{2}_{\textrm{S+A}} = 
\sum_{p=1}^{4} 
\chi_{p\textrm{, }A_\textrm{DN}}^{2}
+ \chi_{\textrm{spec}}^{2}.
\end{equation}

While this simpler method accurately predicts the same asymmetries for each oscillation parameter set as the original maximum likelihood method in the SI case, it is less constraining to the solar oscillation parameters due to the loss of the energy and zenith dependence in the predictions, though this information is still present in the recoil electron data.
As a consequence, the so-called LOW solar neutrino solution in the $(\sin^{2}\theta_{12}, \Delta m_{21}^{2})$ parameter space in the SI case that is normally excluded by SK's measured zenith spectrum is no longer excluded at greater than $3\sigma$ by SK data alone. 

However, the SI LOW solution is also excluded by the very small day-night asymmetry in the 862 keV $^{7}$Be solar neutrino flux measurement by Borexino neutrino-electron ES in liquid scintillator: $A_\textrm{DN}^\textrm{Be7} = (-0.1\pm1.2\pm0.7)\%$ \cite{borexino_be7_adn}.
It is worth noting that the day-night asymmetry definition $A_{dn}$ in the Borexino paper is opposite in sign than that of the definition used by SK and adopted in this paper ($A_\textrm{DN} =  -A_{dn}$).
This analysis incorporates a fit to their measurement in order to remove the LOW solution (in the SI case), and the equivalent of the LOW solution with NSI effects implemented, making the analysis more robust. For the case of SI, this procedure excludes most of the LOW solution at greater than 5$\sigma$. 
While $^{7}$Be neutrinos are lower in energy than $^{8}$B and $hep$ neutrinos, their profile of neutrino production in the Sun is  similar to that of the $^{8}$B neutrinos, since the $^{7}$Be neutrinos are produced in the preceding reaction of the branch of the $pp$ fusion chain that produces $^{8}$B neutrinos.
This means that $^{7}$Be neutrinos can test a similar density profile in the Sun as $^{8}$B neutrinos, albeit at a lower energy outside the MSW resonance for the SI case.

In order to reproduce Borexino's $A_\textrm{DN}^\textrm{Be7}$ result, the zenith dependent exposure in Fig. 1 in \cite{borexino_be7_adn} is extracted in order to account for the zenith dependence of their measurement for calculating $P_{ee}^{\textrm{NIGHT}}$. 
The mono-energetic $^{7}$Be neutrino probabilities during the day and zenith directions are calculated using the same procedure as described earlier, and the night probability is calculated based on the zenith probabilities weighted by Borexino's exposure.
In order to approximate the 862 keV $^{7}$Be rate $R$ at Borexino, mono-energetic neutrinos are used. 
The differential rate with respect to $T_{e}$ is then proportional to the combination of survival probabilities and differential cross sections
\begin{equation}
\frac{dR}{dT_{e}} 
\propto  
\bigg[  P^{\textrm{Z}}_{ee}\frac{d\sigma_{\nu_{e}}}{dT_{e}} 
+ (1-P_{ee}^{\textrm{Z}})\frac{d\sigma_{\nu_{\beta}}}{dT_{e}}\bigg] ,
\end{equation}
where the cross sections $\sigma$ and $P_{ee}$ have the neutrino energy fixed by the Dirac delta function and  $\nu_{\beta}$ corresponds to the contribution from both $\nu_{\mu}$ and $\nu_{\tau}$.
Here, again, Z = D (Z) for day (night).
Since the Borexino analysis sets the range of their observed recoil electron kinetic energy between 550 and 800 keV, we integrate the cross sections over $T_{e}$ up to the maximum allowed by conservation of four momentum ($T_{e}^{\textrm{max}} \sim 665$ keV).
These neutrino-electron elastic scattering cross sections include radiative corrections described in \cite{bahcall_hep_cx}.
When forming the day-night asymmetry for 862 keV $^{7}$Be after integrating over $T_{e}$, the flux and target information cancel, and the predicted day-night asymmetry simply becomes
\begin{equation}
A_\textrm{DN}^{Be7} =
2\frac{\sigma_{e}P_{ee}^{\textrm{D}}  + \sigma_{x}(1-P_{ee}^{\textrm{D}}) - [\sigma_{e}P_{ee}^{\textrm{N}}  + \sigma_{x}(1-P_{ee}^{\textrm{N}}) ]}
{\sigma_{e}P_{ee}^{\textrm{D}}  + \sigma_{x}(1-P_{ee}^{\textrm{D}}) + \sigma_{e}P_{ee}^{\textrm{N}}  + \sigma_{x}(1-P_{ee}^{\textrm{N}}) }
\end{equation}
where $\sigma_{e(x)} = \sigma_{\nu_{e}(\nu_{x})}$.
A simple $\chi^{2}$ fit of the predicted $^{7}$Be asymmetry to the Borexino measurement is performed, denoted as $\chi^{2}_{\textrm{BBe7A}}$, and the $3\sigma$ exclusion contours reported by the Borexino collaboration in Fig. 5 of \cite{borexino_be7_adn} were accurately reproduced.
As correlations in the flux between the $^{7}$Be and the $^{8}$B (and $hep$) from the SSM is also canceled when forming the day-night asymmetry, $\chi^{2}_\textrm{BBe7A}$ can simply be added to the $\chi^{2}$ from the SK oscillation analysis.
In the SI case, after combing $\chi^{2}$'s with the spectrum and day-night asymmetry fit $\chi^{2}_{\textrm{S+A}}$ to the SK data, the combined $\chi^{2}_{\textrm{comb}}$ is 
\begin{equation}
\label{total_chi2}
\chi^{2}_{\textrm{comb}} =  \chi^{2}_\textrm{S+A} + \chi^{2}_\textrm{BBe7A},
\end{equation}
and the LOW solution is excluded while leaving the $\chi^{2}_{\textrm{comb}}$ in the LMA region unaltered from the $\chi^{2}_{\textrm{S+A}}$ value.
Once including NSI in the matter potential, the $\chi^{2}_{\textrm{BBe7A}}$ removes any LOW-equivalent solutions that return due to the change in the method of including SK $A_\textrm{DN}$ measurements into the NSI analysis. 
When including the data from SNO, the same procedure for combining $\chi^{2}$ applies, and, in the SI case, the LOW solution is excluded while leaving the $\chi^{2}$(SK+SNO) unchanged in the LMA region.
\section{\label{sec:results}Results}
Two sets of results of the SK solar neutrino NSI analysis for up and down quarks will be given here.
The first set of results corresponds to the fit to the spectra and $A_\textrm{DN}$ of all SK phases using $\chi^{2}_{\textrm{S+A}}$ (denoted as SK-Only).
The second set corresponds to the combined results of SK with SNO and Borexino's $^{7}$Be day-night asymmetry (denoted as Combined).
The parameter estimation results will be presented as the two dimensional contours for the 1, 2, and 3$\sigma$ allowed regions for the solar neutrino oscillation parameters $(\sin^{2}\theta_{12},\Delta m_{21}^{2})$ and for the effective NSI parameters $(\epsilon_{11}^{f},\epsilon_{12}^{f})$ slices of the parameter space (where $f = u, d$).
For two degrees of freedom, the 1, 2, and 3$\sigma$ values correspond to $\Delta\chi^2$ values of approximately 2.30, 6.18, and 11.83 respectively.
These contours are formed after profiling over the other parameters not displayed.
For each pair of displayed parameters, profiling over the non-displayed parameters involves choosing the minimum $\chi^2$ value from the corresponding set of non-displayed parameters. 

\begin{figure*}
\center
\begin{tabular}{c c}
\includegraphics[width = 0.48\linewidth]{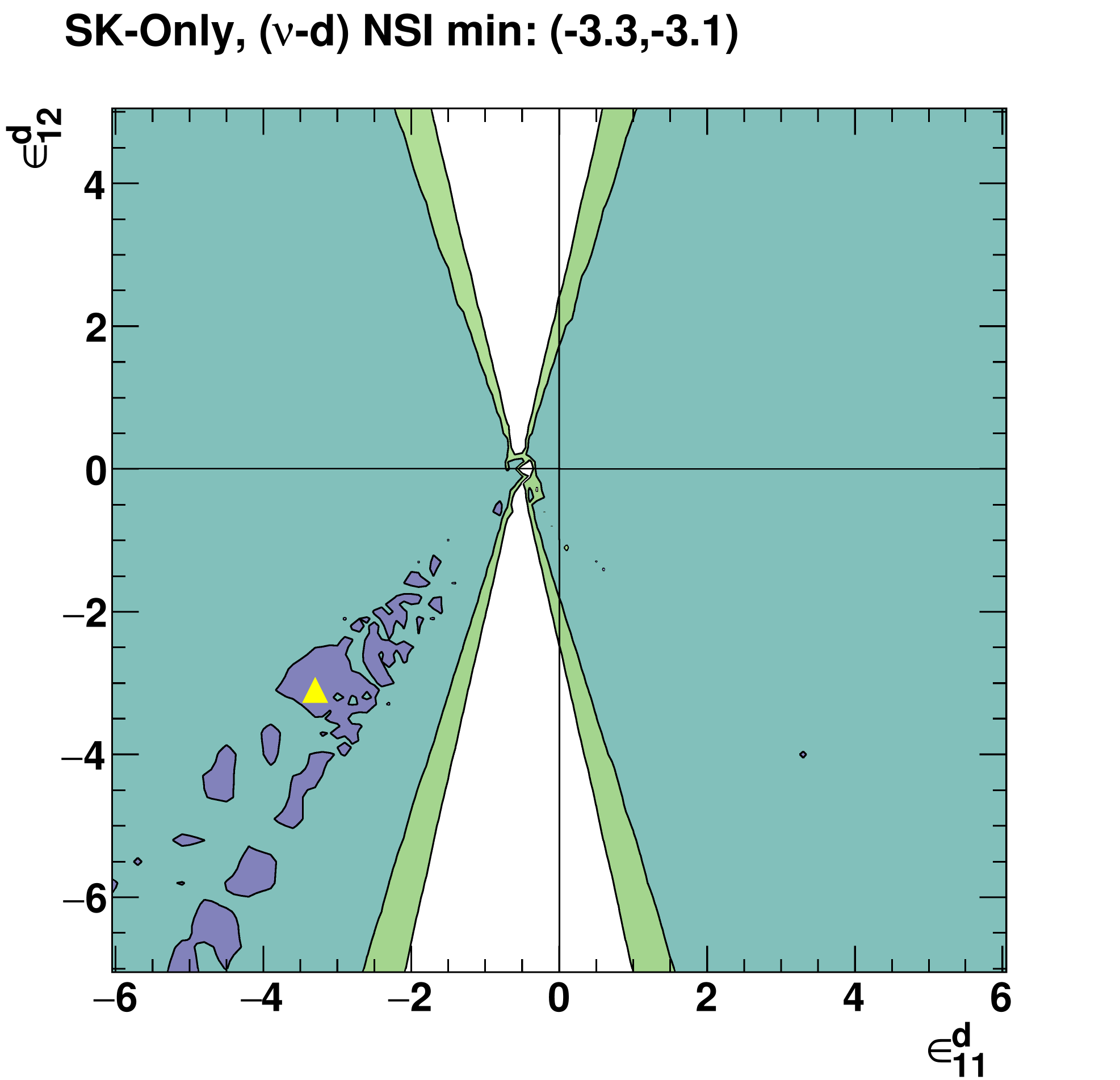} &
\includegraphics[width = 0.48\linewidth]{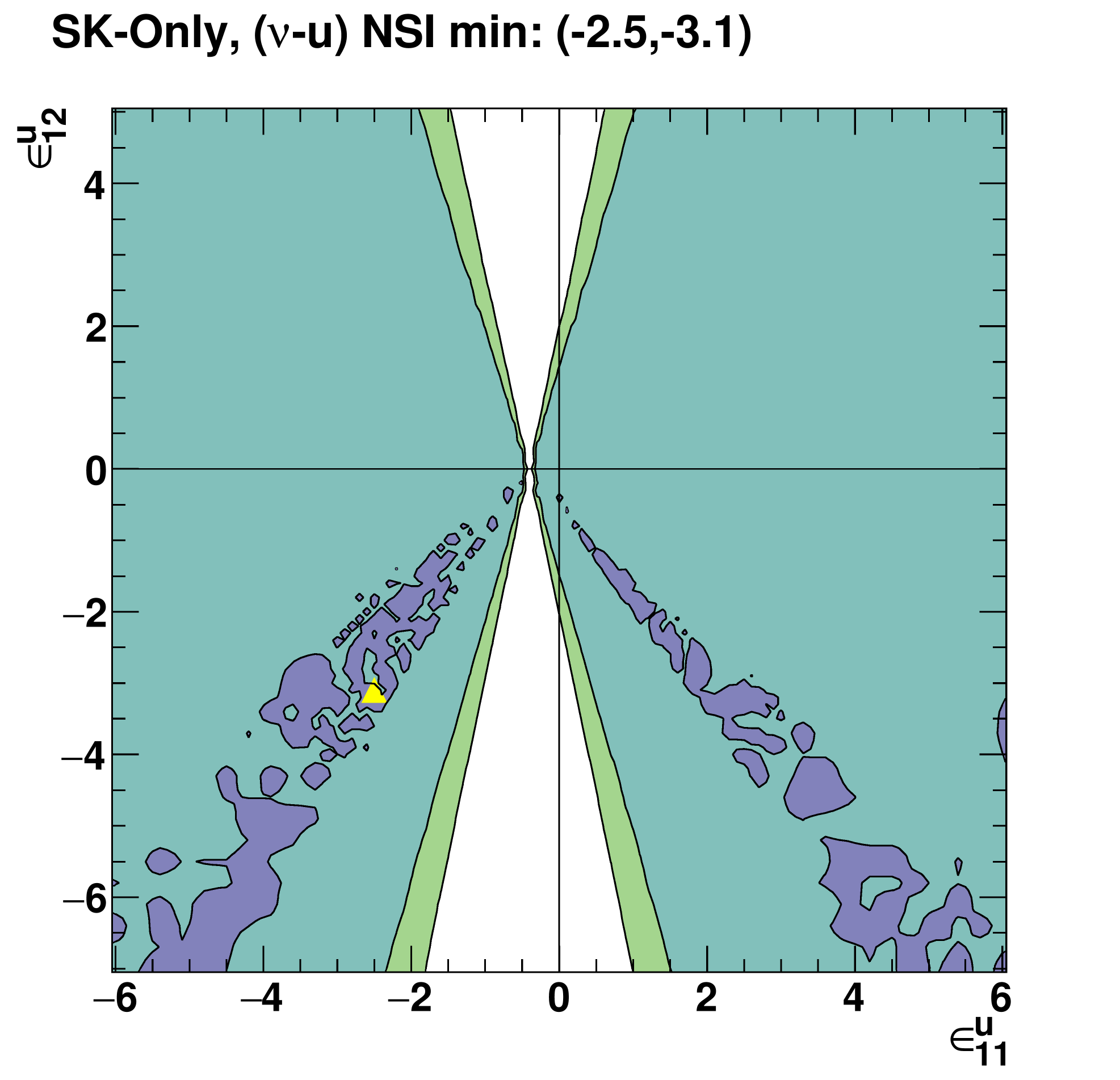} \\
\includegraphics[width = 0.48\linewidth]{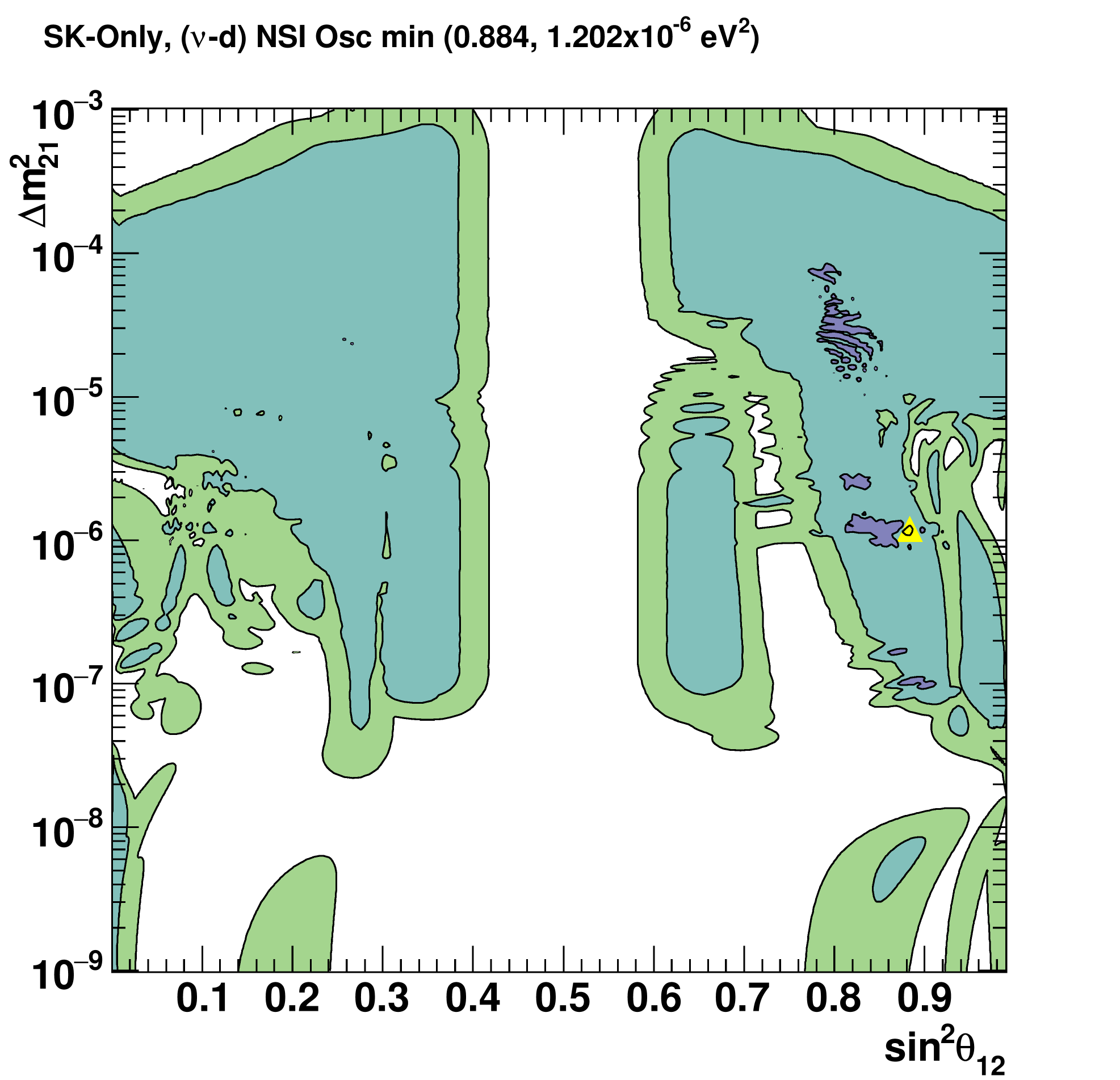} &
\includegraphics[width = 0.48\linewidth]{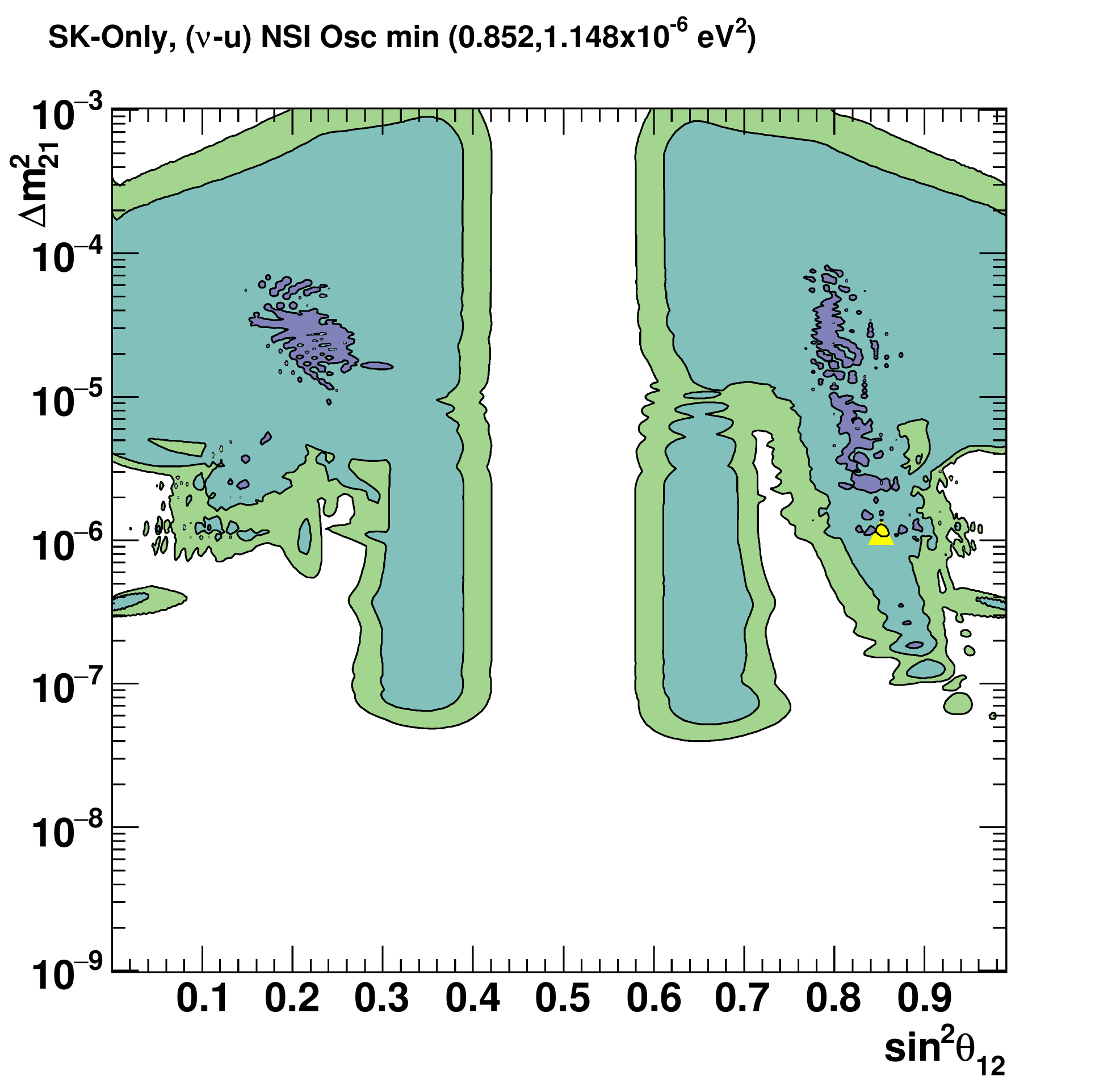} \\

\end{tabular}
\caption{The allowed regions given by the SK-Only fit for one (blue), two (teal), and three (green) $\sigma$ from two-dimensional parameter estimation ($\Delta\chi^2$ with 2 d.o.f).
The best-fit point is denoted by a yellow triangle and written in the title for each plot.
The left column corresponds to ($\nu$-$d$) NSI and the right column corresponds to ($\nu$-$u$) NSI.
The top row corresponds to the effective NSI parameter space ($\epsilon_{11}^f, \epsilon_{12}^f$), while the bottom row corresponds to the SI oscillation parameters ($\sin^2\theta_{12},\Delta m_{21}^2$).
The non-displayed parameters have been profiled over in each plot.}
\label{sk_only_results}
\end{figure*}

For the ($\nu$-$d$) NSI SK-Only case, the $(\epsilon_{11}^{d},\epsilon_{12}^{d})$ contours are given in the top left of Fig. \ref{sk_only_results}. 
The $(\epsilon_{11}^{u},\epsilon_{12}^{u})$ parameter estimation contours from the ($\nu$-$u$) NSI SK-Only fit are provided in the top right of the figure. 
The $(\sin^{2}\theta_{12},\Delta m_{21}^{2})$ parameter estimation contours for ($\nu$-$d$) NSI and ($\nu$-$u$) NSI are given in the bottom right and left of Fig. \ref{sk_only_results} respectively.
In the figures, the 1, 2, and 3$\sigma$ allowed regions are shaded in blue, teal, and green respectively. 
The minimum is denoted by the yellow triangle.

When fitting to the SK-Only data, the best-fit points for the scenarios with ($\nu$-$d$) NSI or ($\nu$-$u$) NSI lie in a LOW-like solution in the ``darkside'' region ($\sin^2\theta_{12} > 0.5$). 
By combining the fit to the SK data with the fit to the Borexino $A_\textrm{DN}^{\textrm{Be7}}$, LOW-like solutions are removed, and the best-fit point shifts to an LMA-like solution.
In the ($\nu$-$u$) NSI scenario, the best-fit point also shifts to the ``lightside'' region ($\sin^2\theta_{12} < 0.5$), while in the ($\nu$-d) NSI scenario, the best-fit point remains in the darkside region.

\begin{figure*}
\center
\begin{tabular}{c}
\includegraphics[width = 0.99\linewidth]{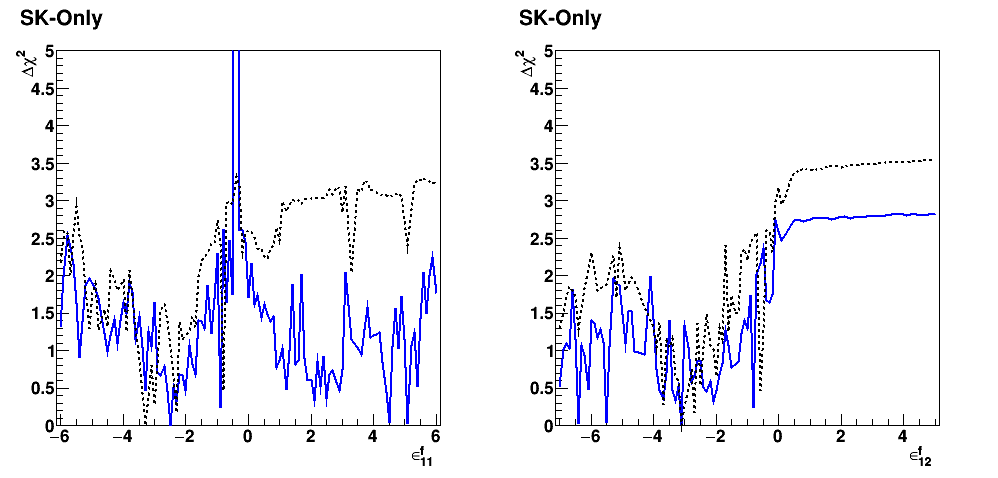} \\ 
\end{tabular}
\caption{The SK-Only one-dimensional $\Delta\chi^2$ limits for the effective NSI parameters $\epsilon_{11}^f$ and $\epsilon_{12}^f$ for $f = u$ or $d$. 
The results for $\epsilon_{11}^f$ ($\epsilon_{12}^f$) are plotted in the left (right) column. 
The solid blue and black-dashed lines correspond to the results for ($\nu$-$u$) NSI and ($\nu$-$d$) NSI respectively.
All other parameters have been profiled over.}
\label{sk_only_1d_results}
\end{figure*}

The one-dimensional $\Delta\chi^2$ limits for the effective NSI parameters are plotted in Fig \ref{sk_only_1d_results}. 
The results for $\epsilon_{11}^f$ ($\epsilon_{12}^f$) are plotted in the left (right). 
The solid blue and black-dashed lines correspond to the results for ($\nu$-$u$) NSI and ($\nu$-$d$) NSI respectively.
The y-axis range has been restricted to $\Delta\chi^2 < 5$ since SK is not so sensitive across most of the range.
While SK data alone only disfavors $\epsilon_{11}^{u}\simeq-0.4$ at $\Delta\chi^2>4$, it also slightly disfavors $\epsilon_{11}^{d} > 0$ at $\Delta\chi^2 > 2$.
SK data additionally disfavors values of $\epsilon_{12}^{u}$ and $\epsilon_{12}^{d}$ above 0.4 at $\Delta\chi^2 > 2.7$ and $\Delta\chi^2 > 3.4$.

\begin{figure*}[]
\includegraphics[width = 0.99\linewidth]{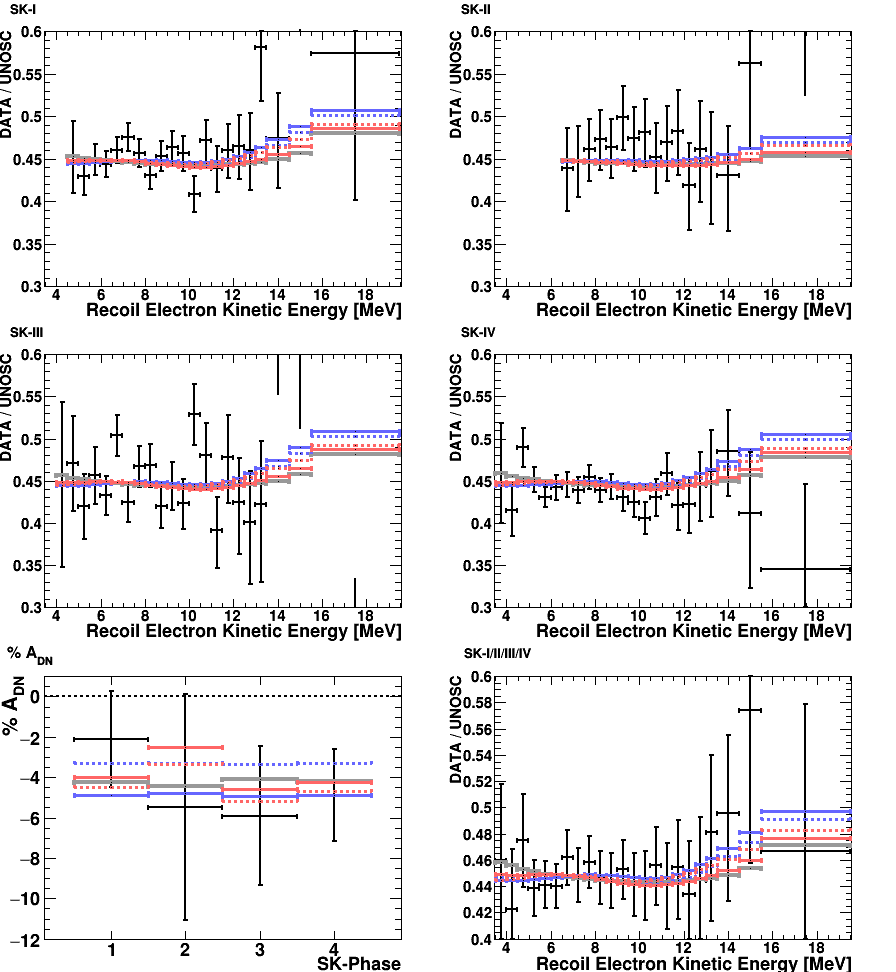}
\caption{The best-fit SK spectral and $A_\textrm{DN}$ predictions from the SK-Only fit (blue) and SK + Borexino $A_{\textrm{DN}}^{ \textrm{Be7} }$ (red).  
The top four panels are the recoil electron spectra for each SK phase. 
The bottom left panel is the day-night asymmetry for each SK phase.
The bottom right panel is the combined SK-I/II/III/IV spectrum, which is provided for illustrative purposes.
The data (black lines) includes statistical and energy-uncorrelated errors added in quadrature.
The solid (dashed) lines correspond to NSI with up quarks (down quarks).
 The LMA best fit for SI (solid gray line) is plotted for comparison.}
\label{fig_skonly_spec} 
\end{figure*}

\begin{figure*}[]
\includegraphics[width = 0.99\linewidth]{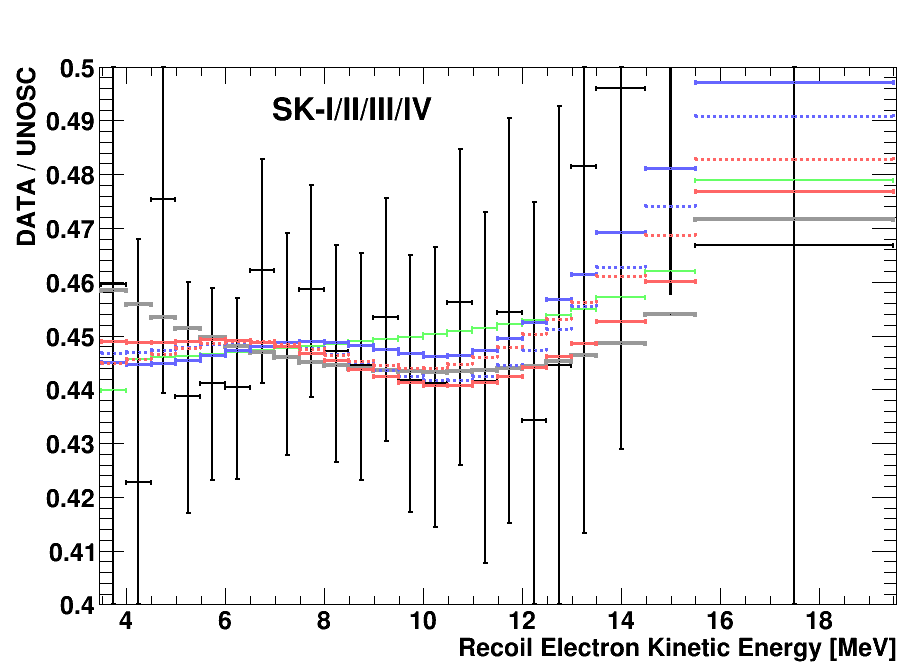}
\caption{The combined SK-I/II/III/IV spectrum from the NSI SK-Only fit (blue) and SK + Borexino $A_{\textrm{DN}}^{\textrm{Be7} }$ (red).
This plot is for illustrative purposes only and should not be used in any analyses.
The data (black lines) includes statistical and energy-uncorrelated errors added in quadrature.
The solid (dashed) lines correspond to NSI with up quarks (down quarks).
 The LMA best fit for SI (solid gray line) and for a constant value of $P_{ee}$ = 0.317 (green) are plotted for comparison.}
\label{sk1234_spectra_skonly_nsi} 
\end{figure*}

In addition to the allowed regions and one-dimensional limits from the parameter estimation, the SK-Only best-fit predictions of the SK recoil electron spectra and $A_\textrm{DN}$ are shown in Fig. \ref{fig_skonly_spec} for each SK phase as blue lines.
The data (black lines) in the figure includes statistical and energy-uncorrelated errors added in quadrature.
The solid (dashed) lines correspond to NSI with up quarks (down quarks) and the SK-Only best-fit results. 
The SK + Borexino $A_\textrm{DN}^\textrm{Be7}$ best-fit results are given as well (red lines).
 The LMA best-fit point for SI (solid gray line) is plotted for comparison.
The bottom right panel in the figure is the SK-I/II/III/IV combined spectra, which is provided for illustrative purposes and should not be used for analysis. 
The SK-Only best fit with ($\nu$-$u$) NSI prediction has a slight downturn below 6.99 MeV, while the SK+Borexino $A_\textrm{DN}^\textrm{Be7}$ prediction is relatively flat below this energy.
The behavior of the ($\nu$-$d$) NSI best fits is the opposite to that of the ($\nu$-$u$) NSI best fits in both cases.
An enlarged version of the SK-I/II/III/IV combined spectrum plot is provided in Fig. \ref{sk1234_spectra_skonly_nsi}.
The enlarged figure additionally includes the combined spectrum prediction resulting from a constant-value solar neutrino survival probability of $P_{ee} = 0.317$ (green line) for comparison.
The constant-value $P_{ee}$ fit to SK spectral data (including day-night effects) has been reported in table X of \cite{SK4Solar}.

Comparisons between the spectral shapes of the SK-Only NSI results (and SK + Borexino $A_\textrm{DN}^\textrm{Be7}$) with the spectrum due to a constant $P_{ee} = 0.317$ 
have been performed to determine if the central energy region of the spectra or the low energy region is providing a better fit.
The energy bins below 5.49 MeV were removed energy bin by energy bin and the best-fit results and the constant $P_{ee}$ spectral predictions were refit for the best-fit results for SI, ($\nu$-$d$) NSI, and ($\nu$-$u$) NSI scenarios.
In all of the fits for this comparison, the $\beta$-parameter corresponding to the $^{8}$B flux in the unbinned maximum likelihood fit is not constrained and the $hep$ flux parameter $\eta$ is weakly constrained in order to test the spectral shapes.
The results of the refits to the SK spectra with the removal of lower energy spectral data are reported in Table \ref{compare_flat_fit_table}. 
The table contains both SK-Only and SK + Borexino $A_\textrm{DN}^{\textrm{Be7}}$ refits, as well as the original full spectrum best-fit results for comparison.
As the lower energy bins were removed from the fit, the SK-I/II/III/IV combined spectral fit was consistently better for the NSI scenarios than with either the constant-value $P_{ee}$ case or either SI cases (LMA or LOW). 
When comparing each SK experimental phase, the NSI scenarios better fit the spectral shape of all phases than the constant-value $P_{ee}$, with SK-IV and SK-I significantly contributing to the improved fit.
The SI best fits (LOW and LMA) of the combined SK-I/II/III/IV spectral fit were also consistently better than the constant-value $P_{ee}$ predictions.
The better spectral shape agreement in the SI best-fit cases compared to the constant-value $P_{ee}$ prediction comes from the lower energy bins of the SK-IV spectrum and the central energy bin regions for SK-I, SK-III, and SK-IV.
This indicates that the central energy region of the spectral fits, specifically for SK-I and SK-IV, are the dominant contribution to the better $\chi^2$ from the spectral fit.
The same comparison procedure to the flat $P_{ee}$ prediction used in Table \ref{compare_flat_fit_table} was performed for the SK+SNO+Borexino $A_\textrm{DN}^\textrm{Be7}$ combined fit results and are reported in Table \ref{compare_flat_fit_combined_table}.
For the determination of the combined fit for flat $P_{ee}$, the SNO coefficients were determined assuming the flat $P_{ee}$ as the daytime $^8$B $P_{ee}$,
and the $\eta$ $hep$ flux parameter is constrained for the range over which the use of the SNO coefficients are valid.
Similar to the comparison between the flat $P_{ee}$ SK spectral fits and the SI and NSI spectral fits in Talbe \ref{compare_flat_fit_table},
the comparisons between the combined fits and the combined fit for the flat $P_{ee}$ show a similar trend. 
However, the $\chi^{2}$ reduction when removing lower energy bins is smaller than when fitting only to the SK spectra.
This effect is due to the lower energy component of the SNO data remaining in the combined fit, which cannot be changed due to the method of including the SNO data using the SNO coefficients.

The comparison between the best-fit points with and without NSI are performed using a one degree of freedom log-likelihood ratio (LLR):
\begin{equation} 
\textrm{LLR} = \log\mathcal{L}\textrm{(NSI-best)} - \log\mathcal{L}\textrm{(SI-best)}.
\label{llr_eq}
\end{equation}
A log-likelihood ratio is used to determine the $\sigma$ value since the errors of the spectral bins are approximately Gaussian.
The LLR for the SK-Only fit to the ($\nu$-$u$) NSI scenario is 1.33, with a corresponding sigma-value of 1.63.
In the ($\nu$-$d$) NSI scenario, the  LLR for the SK-Only fit is 1.63 with a sigma-value of 1.80.
With the two additional parameters contributing to the shape of the $P_{ee}$ and therefore adding two additional degrees of freedom to the fit, 
the LLR is expected to be approximately 1.0, and the sigma-value is expected to be $\sim$1.4.
The results show a slightly stronger preference for NSI to SI than is expected by the increased degrees of freedom in the fit, though it is still only a slight preference.
 
\begin{table*}[]
\caption{The comparisons between the fits to the SK spectra from a constant $P_{ee} = 0.317$ and the best-fit spectra from the SI case (LMA and LOW), and between the constant $P_{ee}$ spectra and the best-fit spectra with NSI effects (SK-Only and SK + Borexino $A_\textrm{DN}^{\textrm{Be7}}$).
The first table contains the results for fitting to the full spectra. 
The second table is the results of fitting to the spectra above 5.49 MeV. 
The difference between the constant value $P_{ee}$ spectral fit ($\chi^2_\textrm{cv}$) and the spectral fit $\chi^2$ ($\chi^2_\textrm{spec}$) is given as $\Delta\chi^2_\textrm{cv} = \chi^2_\textrm{cv} - \chi^2_\textrm{spec}$.
Better fits will have positive values of $\Delta\chi^2_\textrm{cv}$, and poorer fits will have negative values.
Dashes indicate fits that are unaffected by changing the lower threshold of the spectral fit.
For the 4.49 MeV threshold, three spectral data points are removed from the fit: one from SK-III and two from SK-IV.
By increasing the threshold to 5.49 MeV, nine spectral data points are removed from the fit: two, four, and five data points from SK-I, SK-III, and SK-IV respectively.
}
\label{compare_flat_fit_table}
\center
\begin{tabular*}{\textwidth}{l @{\extracolsep{\fill}} c c c c c}
\hline 
\hline
Prediction  &   SK-I/II/III/IV & SK-I & SK-II & SK-III & SK-IV (1664 days)\\
\hline 
 & & & & & \\
Threshold = 3.49 MeV & & & & & \\
Constant $P_{ee}$ SK-Only Fit & 69.30 &18.92	& 5.30	& 27.94 &	15.5\\
& $\begin{matrix}\chi^2\textrm{  }&\Delta\chi^2_\textrm{cv}\end{matrix}$ & $\begin{matrix}\chi^2\textrm{  }&\Delta\chi^2_\textrm{cv}\end{matrix}$& $\begin{matrix}\chi^2\textrm{  }&\Delta\chi^2_\textrm{cv}\end{matrix}$&$\begin{matrix}\chi^2\textrm{  }&\Delta\chi^2_\textrm{cv}\end{matrix}$&$\begin{matrix}\chi^2\textrm{  }&\Delta\chi^2_\textrm{cv}\end{matrix}$\\
SI LOW & $\begin{matrix} 68.26 &1.04\end{matrix}$& $\begin{matrix} 19.32 &-0.40\end{matrix}$	
&$\begin{matrix}5.45 &-0.15\end{matrix}$ &$\begin{matrix}27.97 &-0.03\end{matrix}$ & $\begin{matrix}13.97 &1.54\end{matrix}$\\
SI LMA & $\begin{matrix}68.38 &0.92 \end{matrix}$&$\begin{matrix}19.39 &-0.47 \end{matrix}$&$\begin{matrix}5.43 &-0.13 \end{matrix}$&
$\begin{matrix}28.14 &-0.20 \end{matrix}$&$\begin{matrix}13.79 & 1.71 \end{matrix}$\\
($\nu$-$d$) SK-Only &$\begin{matrix} 64.14 & 5.16\end{matrix}$&$\begin{matrix}17.29 & 1.63\end{matrix}$&
$\begin{matrix}5.08 & 0.22  \end{matrix}$&$\begin{matrix}27.32 & 0.62  \end{matrix}$&
$\begin{matrix}12.93 & 2.57 \end{matrix}$\\
($\nu$-$d$) SK+Borexino $A_\textrm{DN}^{\textrm{Be7}}$&$\begin{matrix}65.36 &3.94 \end{matrix}$&$\begin{matrix}17.82 &1.10 \end{matrix}$&
$\begin{matrix}5.27 &0.03 \end{matrix}$&$\begin{matrix}27.46 &0.48 \end{matrix}$&
$\begin{matrix} 13.40 &2.10\end{matrix}$\\
($\nu$-$u$) SK-Only &$\begin{matrix}64.61 &4.69 \end{matrix}$&$\begin{matrix}17.40 &1.52 \end{matrix}$&
$\begin{matrix}5.07 &0.23 \end{matrix}$&$\begin{matrix}27.18 &0.76 \end{matrix}$&
$\begin{matrix}13.61 &1.89 \end{matrix}$\\
($\nu$-$u$) SK+Borexino $A_\textrm{DN}^{\textrm{Be7}}$&$\begin{matrix}65.71&3.59\end{matrix}$&$\begin{matrix}18.06&0.87\end{matrix}$&$\begin{matrix}5.24&0.06\end{matrix}$&
$\begin{matrix}27.75&0.19\end{matrix}$&$\begin{matrix}13.13&2.37\end{matrix}$\\
\hline 
 & & & & & \\
Threshold = 4.49 MeV & & & & & \\
Constant $P_{ee}$ SK-Only Fit & 68.16 & $-$ & $-$ &27.94	&14.31\\
& $\begin{matrix}\chi^2&\Delta\chi^2_\textrm{cv}\end{matrix}$ & & &$\begin{matrix}\chi^2&\Delta\chi^2_\textrm{cv}\end{matrix}$&$\begin{matrix}\chi^2&\Delta\chi^2_\textrm{cv}\end{matrix}$\\
SI LOW & $\begin{matrix}66.76 &1.40 \end{matrix}$& $-$ & $-$ &
$\begin{matrix}27.96 &-0.02 \end{matrix}$&$\begin{matrix}12.68  &1.63 \end{matrix}$\\
SI LMA & $\begin{matrix}66.74 &1.42 \end{matrix}$& $-$ & $-$ &
$\begin{matrix}28.12 &-0.19 \end{matrix}$&$\begin{matrix}12.39 &1.91 \end{matrix}$\\
($\nu$-$d$) SK-Only & $\begin{matrix}63.08 &5.07 \end{matrix}$& $-$ & $-$ &
$\begin{matrix}27.32 &0.62 \end{matrix}$&$\begin{matrix}12.01 &2.30 \end{matrix}$\\
($\nu$-$d$) SK+Borexino $A_\textrm{DN}^{\textrm{Be7}}$& $\begin{matrix}64.36 &3.80 \end{matrix}$& $-$ & $-$ &
$\begin{matrix}27.46 &0.48 \end{matrix}$&$\begin{matrix}12.53 &1.78 \end{matrix}$\\
($\nu$-$u$) SK-Only &$\begin{matrix}63.66 &4.50 \end{matrix}$& $-$ & $-$ &
$\begin{matrix}27.18 &0.76 \end{matrix}$&$\begin{matrix}12.77 &1.53 \end{matrix}$\\
($\nu$-$u$) SK+Borexino $A_\textrm{DN}^{\textrm{Be7}}$& $\begin{matrix}64.55 &3.61 \end{matrix}$& $-$ & $-$ &
$\begin{matrix}27.74 &0.19 \end{matrix}$&$\begin{matrix}12.14 &2.17 \end{matrix}$\\
\hline 
 & & & & & \\
Threshold = 5.49 MeV & & & & & \\
Constant $P_{ee}$ SK-Only Fit &62.50&18.73& $-$ &27.17&9.02\\  
& $\begin{matrix}\chi^2&\Delta\chi^2_\textrm{cv}\end{matrix}$ & $\begin{matrix}\chi^2&\Delta\chi^2_\textrm{cv}\end{matrix}$&& 
$\begin{matrix}\chi^2&\Delta\chi^2_\textrm{cv}\end{matrix}$&$\begin{matrix}\chi^2&\Delta\chi^2_\textrm{cv}\end{matrix}$\\
SI LOW &$\begin{matrix}61.98&0.51\end{matrix}$&$\begin{matrix}18.26&0.47\end{matrix}$& $-$ &
$\begin{matrix}27.02&0.15\end{matrix}$&$\begin{matrix}8.90&0.12 \end{matrix}$\\ 
SI LMA &$\begin{matrix}62.05 &0.44 \end{matrix}$&$\begin{matrix}18.21&0.52 \end{matrix}$& $-$ &
$\begin{matrix}27.12 &0.05 \end{matrix}$&$\begin{matrix}8.90&0.13 \end{matrix}$\\
($\nu$-$d$) SK-Only& $\begin{matrix}57.69&4.80\end{matrix}$&$\begin{matrix}16.45 &2.28 \end{matrix}$& $-$ &
$\begin{matrix}26.45&0.71\end{matrix}$&$\begin{matrix}7.28 &1.74\end{matrix}$\\
($\nu$-$d$) SK+Borexino $A_\textrm{DN}^{\textrm{Be7}}$&$\begin{matrix}58.90&3.59\end{matrix}$&$\begin{matrix}17.02&1.71\end{matrix}$& $-$ &
$\begin{matrix}26.61&0.55\end{matrix}$&$\begin{matrix}7.69&1.33\end{matrix}$\\
($\nu$-$u$) SK-Only & $\begin{matrix}57.90&4.59\end{matrix}$&$\begin{matrix}16.74&1.99\end{matrix}$& $-$ &
$\begin{matrix}26.41&0.75\end{matrix}$&$\begin{matrix}7.39&1.63\end{matrix}$\\
($\nu$-$u$) SK+Borexino $A_\textrm{DN}^{\textrm{Be7}}$&$\begin{matrix}59.40&3.10\end{matrix}$&$\begin{matrix}17.15 &1.58\end{matrix}$& $-$ &
$\begin{matrix}26.84&0.33\end{matrix}$&$\begin{matrix}7.76&1.26\end{matrix}$\\
\hline 
\hline
\end{tabular*}
\end{table*} 
\begin{table*}[]
\caption{The comparisons between the combined fits to the SK spectra and SNO coefficients from a constant $P_{ee}=0.317$ and the combined best-fit of the SK, SNO and Borexino $A_\textrm{DN}^{\textrm{Be7}}$ (combined) data for the SI case and with NSI effects.
The first table contains the results for fitting to the full SK spectra. 
The second table is the results of fitting to the SK spectra above 5.49 MeV. 
The difference between the constant value $P_{ee}$ spectral fit ($\chi^2_\textrm{cv}$) and the spectral fit $\chi^2$ ($\chi^2_\textrm{spec}$) is given as $\Delta\chi^2_\textrm{cv} = \chi^2_\textrm{cv} - \chi^2_\textrm{combined}$.
Better fits will have positive values of $\Delta\chi^2_\textrm{cv}$, and poorer fits will have negative values.
Dashes indicate fits that are unaffected by changing the lower threshold of the spectral fit.
For the 4.49 MeV threshold, three spectral data points are removed from the fit: one from SK-III and two from SK-IV.
By increasing the threshold to 5.49 MeV, nine spectral data points are removed from the fit: two, four, and five data points from SK-I, SK-III, and SK-IV respectively.
}
\label{compare_flat_fit_combined_table}
\center
\begin{tabular*}{\textwidth}{l @{\extracolsep{\fill}} c c c c c}
\hline 
\hline
Prediction  &   SK-I/II/III/IV & SK-I & SK-II & SK-III & SK-IV (1664 days)\\
\hline 
 & & & & & \\
Threshold = 3.49 MeV & & & & & \\
Constant $P_{ee}$ Combined Fit & 82.87 &27.95	& 12.48	& 37.45 &	24.06\\
& $\begin{matrix}\chi^2\textrm{  }&\Delta\chi^2_\textrm{cv}\end{matrix}$ & $\begin{matrix}\chi^2\textrm{  }&\Delta\chi^2_\textrm{cv}\end{matrix}$& $\begin{matrix}\chi^2\textrm{  }&\Delta\chi^2_\textrm{cv}\end{matrix}$&$\begin{matrix}\chi^2\textrm{  }&\Delta\chi^2_\textrm{cv}\end{matrix}$&$\begin{matrix}\chi^2\textrm{  }&\Delta\chi^2_\textrm{cv}\end{matrix}$\\
SI Combined (LMA) &$\begin{matrix}74.49 &8.37 \end{matrix}$&$\begin{matrix}23.26 &4.69 \end{matrix}$&
$\begin{matrix}8.87 &3.60 \end{matrix}$&$\begin{matrix}32.91 &4.54 \end{matrix}$&
$\begin{matrix} 16.35 &7.71\end{matrix}$\\
($\nu$-$d$) Combined &$\begin{matrix}68.64 &14.23 \end{matrix}$&$\begin{matrix}19.53 &8.42 \end{matrix}$&
$\begin{matrix}6.23 &6.25 \end{matrix}$&$\begin{matrix}28.18 &9.26 \end{matrix}$&
$\begin{matrix} 15.58 &8.48\end{matrix}$\\
($\nu$-$u$) Combined &$\begin{matrix}71.82&11.05\end{matrix}$&$\begin{matrix}22.04&5.91\end{matrix}$&$\begin{matrix}8.71&3.76\end{matrix}$&
$\begin{matrix}32.16&5.28\end{matrix}$&$\begin{matrix}16.29&7.77\end{matrix}$\\
\hline 
 & & & & & \\
Threshold = 4.49 MeV & & & & & \\
Constant $P_{ee}$ Combined Fit & 82.06 &$-$&$-$& 37.43	&23.35\\
& $\begin{matrix}\chi^2&\Delta\chi^2_\textrm{cv}\end{matrix}$ & & &$\begin{matrix}\chi^2&\Delta\chi^2_\textrm{cv}\end{matrix}$&$\begin{matrix}\chi^2&\Delta\chi^2_\textrm{cv}\end{matrix}$\\
SI Combined (LMA) &$\begin{matrix}72.98 &9.08 \end{matrix}$& $-$ & $-$ &$\begin{matrix}32.90 &4.53 \end{matrix}$&
$\begin{matrix} 15.02 &8.32\end{matrix}$\\
($\nu$-$d$) Combined &$\begin{matrix}67.66 &14.40 \end{matrix}$& $-$ & $-$ &$\begin{matrix}28.18 &9.24 \end{matrix}$&
$\begin{matrix} 14.66 &8.69\end{matrix}$\\
($\nu$-$u$) Combined &$\begin{matrix}70.88 &11.18 \end{matrix}$& $-$ & $-$ &$\begin{matrix}32.16 &5.27 \end{matrix}$&
$\begin{matrix} 15.46 &7.89\end{matrix}$\\
\hline 
 & & & & & \\
Threshold = 5.49 MeV & & & & & \\
Constant $P_{ee}$ Combined Fit &74.82&27.61&$-$&36.82&16.09\\  
& $\begin{matrix}\chi^2&\Delta\chi^2_\textrm{cv}\end{matrix}$ & $\begin{matrix}\chi^2&\Delta\chi^2_\textrm{cv}\end{matrix}$&& 
$\begin{matrix}\chi^2&\Delta\chi^2_\textrm{cv}\end{matrix}$&$\begin{matrix}\chi^2&\Delta\chi^2_\textrm{cv}\end{matrix}$\\
SI Combined (LMA) &$\begin{matrix}68.02 &6.80 \end{matrix}$&$\begin{matrix}22.36 &5.25 \end{matrix}$& $-$ &
$\begin{matrix}32.20 &4.63 \end{matrix}$&$\begin{matrix} 11.24 &4.83\end{matrix}$\\
($\nu$-$d$) Combined &$\begin{matrix}61.90 &12.92 \end{matrix}$&$\begin{matrix}19.01 &8.60 \end{matrix}$& $-$ &
$\begin{matrix}27.50 &9.32 \end{matrix}$&$\begin{matrix} 9.75 &6.32\end{matrix}$\\
($\nu$-$u$) Combined &$\begin{matrix}64.90&9.92\end{matrix}$&$\begin{matrix}21.52&6.09\end{matrix}$& $-$ &
$\begin{matrix}31.53&5.29\end{matrix}$&$\begin{matrix}10.04&6.03\end{matrix}$\\
\hline 
\hline
\end{tabular*}
\end{table*}  
\begin{table*}
\caption{The results of the SK solar neutrino NSI analysis. Each line corresponds to the best fit for the given data set stated in the first column. SK-Only denotes the use of the SK recoil electron spectra and integrated $A_\textrm{DN}$ data for SK-I/II/III/IV while using the SNO NC measurement.
The SK-Only result is in an LOW-like solution of the $(\sin^2\theta_{12},\Delta m_{21}^2)$ parameter space.
SK+Borexino denotes the combined $\chi^2$ of SK-Only result and Borexino's measure of the day-night asymmetry for 862 keV $^{7}$Be solar neutrinos.
The SK+Borexino result removes LOW-like solutions and the best fit is in an LMA-like solution.
SK+SNO denotes the combined fit to SK and SNO data.
Combined denotes the combined fit to the SK+SNO, and Borexino $A_\textrm{DN}^{\textrm{Be7}}$ measurements. 
The second column corresponds to the fermion $f$ with which the NSI occur, for the SI case ($-$), or for $u$-quarks $d$-quarks. 
The third and fourth columns are the resulting best-fit NSI parameters 
$\epsilon_{11}^f$ and $\epsilon_{11}^f$ ($f = u$, or $d$) for the fit. 
Similarly, the fifth and sixth columns are the best-fit $\sin^2\theta_{12}$ and $\Delta m_{21}^2$.
The value of $\sin^2\theta_{13} = 0.020$ is fixed for this analysis.
Results with $\Delta m_{21}^2 < 10^{-5}$ eV$^2$ are in LOW-like solutions, while above this value, they are in LMA-like solutions.
 The last column contains the $\chi^2$ for the given best-fit point and the number of degrees of freedom (NDF).
 For SI, there are two degrees of freedom from $\sin^2\theta_{12}$ and $\Delta m_{21}^2$. 
 With the inclusion of NSI, two additional degrees of freedom are added:  $(\epsilon_{11}^f, \epsilon_{12}^f)$, where $f = u$, or $d$. 
  The bottom table contains the log-likelihood ratio (1 d.o.f) of the NSI best fit, the SI best fit, the corresponding $\sigma$-value for SK-Only, SK+Borexino $A_\textrm{DN}^{\textrm{Be7}}$, and the Combined fit of SK+SNO+Borexino $A_\textrm{DN}^{\textrm{Be7}}$.}
\center
\begin{tabular*}{\textwidth}{c @{\extracolsep{\fill}} c c c c c c}
\hline
\hline
$f$-Quark NSI & Fitted Data & $\epsilon_{11}^f$ & $\epsilon_{12}^f$ & $\sin^2\theta_{12}$ & $\Delta m_{21}^2$ [eV$^2$] & $\chi^2$/NDF\\
\hline
\\
$-$ &SK-Only &  0 & 0 & 0.3442 & $1.258\times10^{-7}$ & 69.4/85\\
$-$ &SK+Borexino & 0 & 0 & 0.3339 & $ 4.073\times10^{-5}$ &  69.6/86\\
$-$ &SK+SNO & 0 & 0 & 0.3137  & $ 4.897\times10^{-5}$ & 76.0/90\\
$-$ &Combined &  0 & 0 & 0.3137 & $4.897\times10^{-5}$ & 76.0/91\\ 
($\nu$-$u$) &SK-Only &  $-$2.5 & $-$3.1 & 0.8519 & $1.148\times10^{-6}$ &66.8/83\\  
($\nu$-$u$) & SK+Borexino & 5.1& $-$6.4 & 0.1864 & $5.957\times10^{-5}$& 66.8/84\\ 
($\nu$-$u$) & SK+SNO & $-$2.2 & $-$1.4 & 0.9380 & $1.259\times10^{-6}$& 72.3/88\\ 
($\nu$-$u$) &Combined &  0.5& $-$1.0 & 0.1935& $1.175\times10^{-5}$ &73.0/89\\ 
($\nu$-$d$) &SK-Only &  $-$3.3 & $-$3.1 & 0.8835 & $1.202\times10^{-6}$ & 66.2/83\\ 
($\nu$-$d$) & SK+Borexino & $-$2.4 & $-$2.0 & 0.8340 & $2.113\times10^{-5}$ & 66.8/84\\ 
($\nu$-$d$) & SK+SNO & $-$3.3 & $-$3.1 & 0.8835 & $1.202\times10^{-6}$& 70.6/88\\ 
($\nu$-$d$) &Combined &  $-$5.1 & $-$6.7 & 0.7992& $8.035\times10^{-5}$ &72.2/89\\ 
\hline
\hline
&&&&&&\\
\end{tabular*}
\begin{tabular}{c}
1 D.O.F. Log-Likelihood Ratio (LLR = $\log\mathcal{L}\textrm{(NSI-best)} - \log\mathcal{L}\textrm{(SI-best)}$) and corresponding $\sigma$-value.
\end{tabular}
\begin{tabular*}{\textwidth}{c @{\extracolsep{\fill}} c c c}
\hline
\hline
NSI&& & SK+SNO+Borexino\\
Interaction & SK-Only & SK+Borexino & (Combined Fit)\\
\hline
%
$\begin{matrix}  \\ (\nu\textrm{-}u) \\ (\nu\textrm{-}d) \end{matrix}$ & 
$\begin{matrix} \textrm{LLR}&\sigma \\ 1.3&1.6 \\ 1.6&1.8 \end{matrix}$ &
$\begin{matrix} \textrm{LLR}&\sigma \\ 1.4&1.7 \\ 1.4&1.7 \end{matrix}$ &
$\begin{matrix} \textrm{LLR}&\sigma \\ 1.5&1.8 \\ 1.9&1.9 \end{matrix}$ \\
\hline
\hline
\end{tabular*}
\label{nsi_results_table_large}
\end{table*}


\begin{figure*}
\center
\begin{tabular}{c c}
\includegraphics[width = 0.48\linewidth]{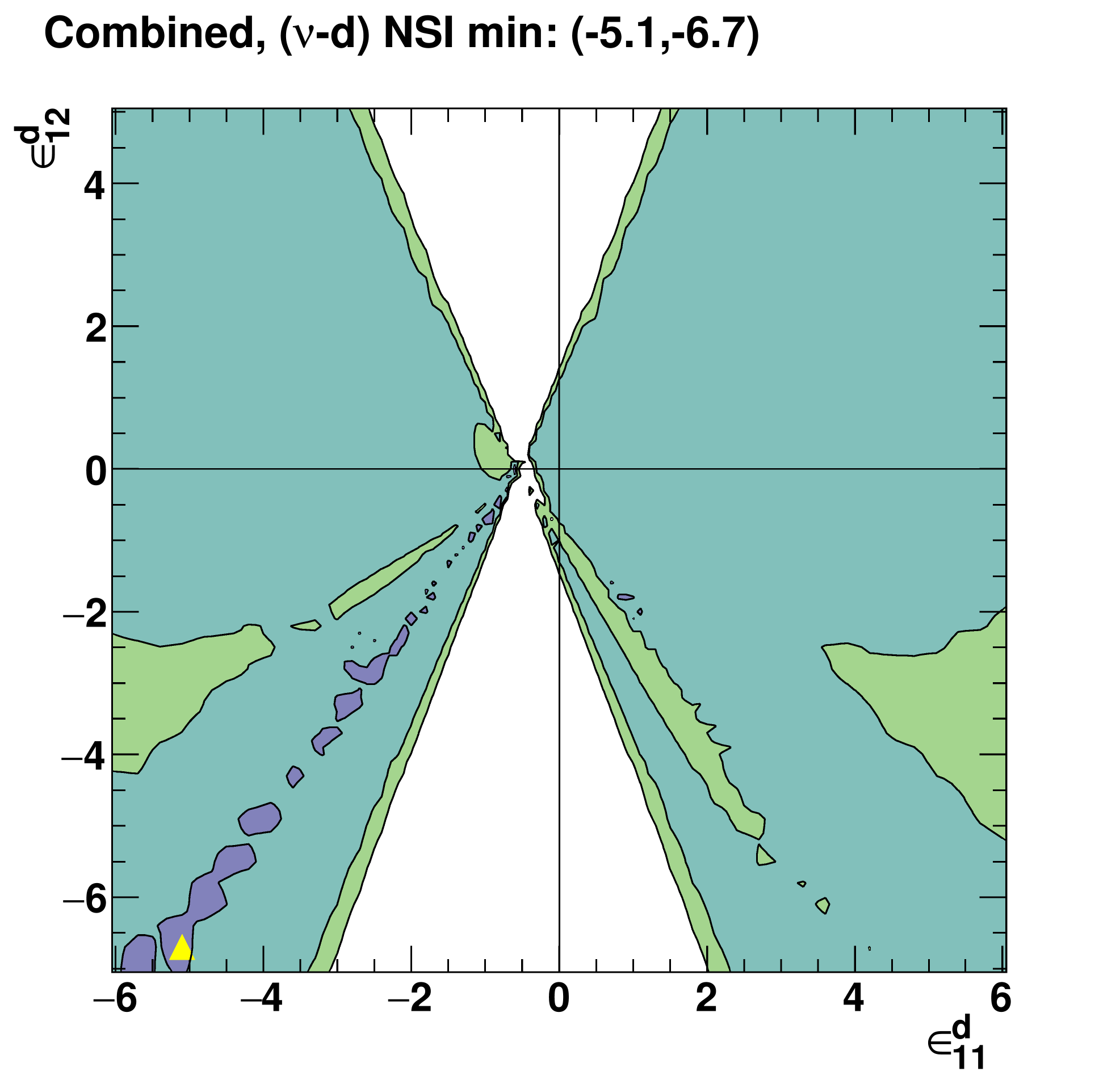} &
\includegraphics[width = 0.48\linewidth]{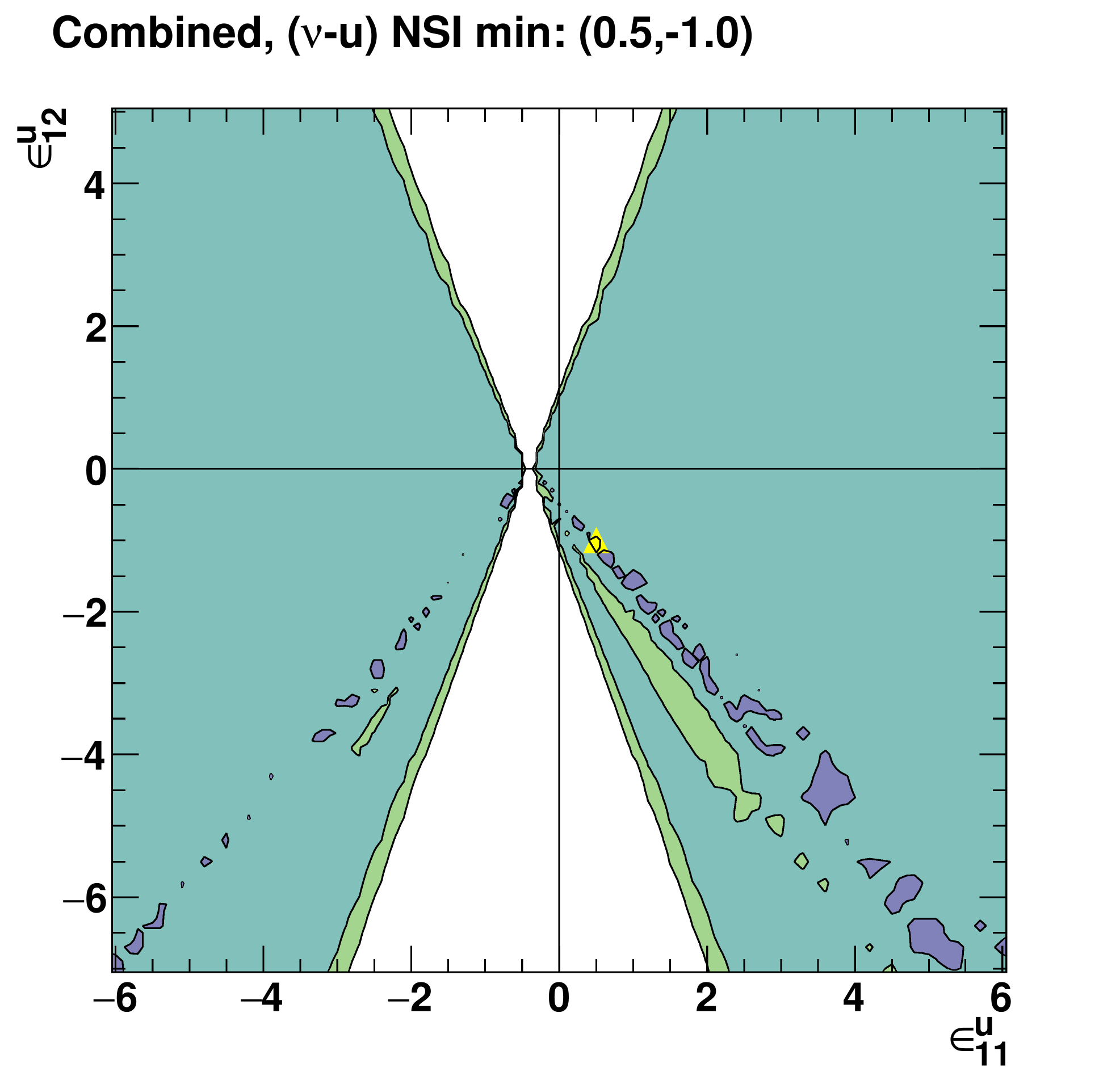} \\
\includegraphics[width = 0.48\linewidth]{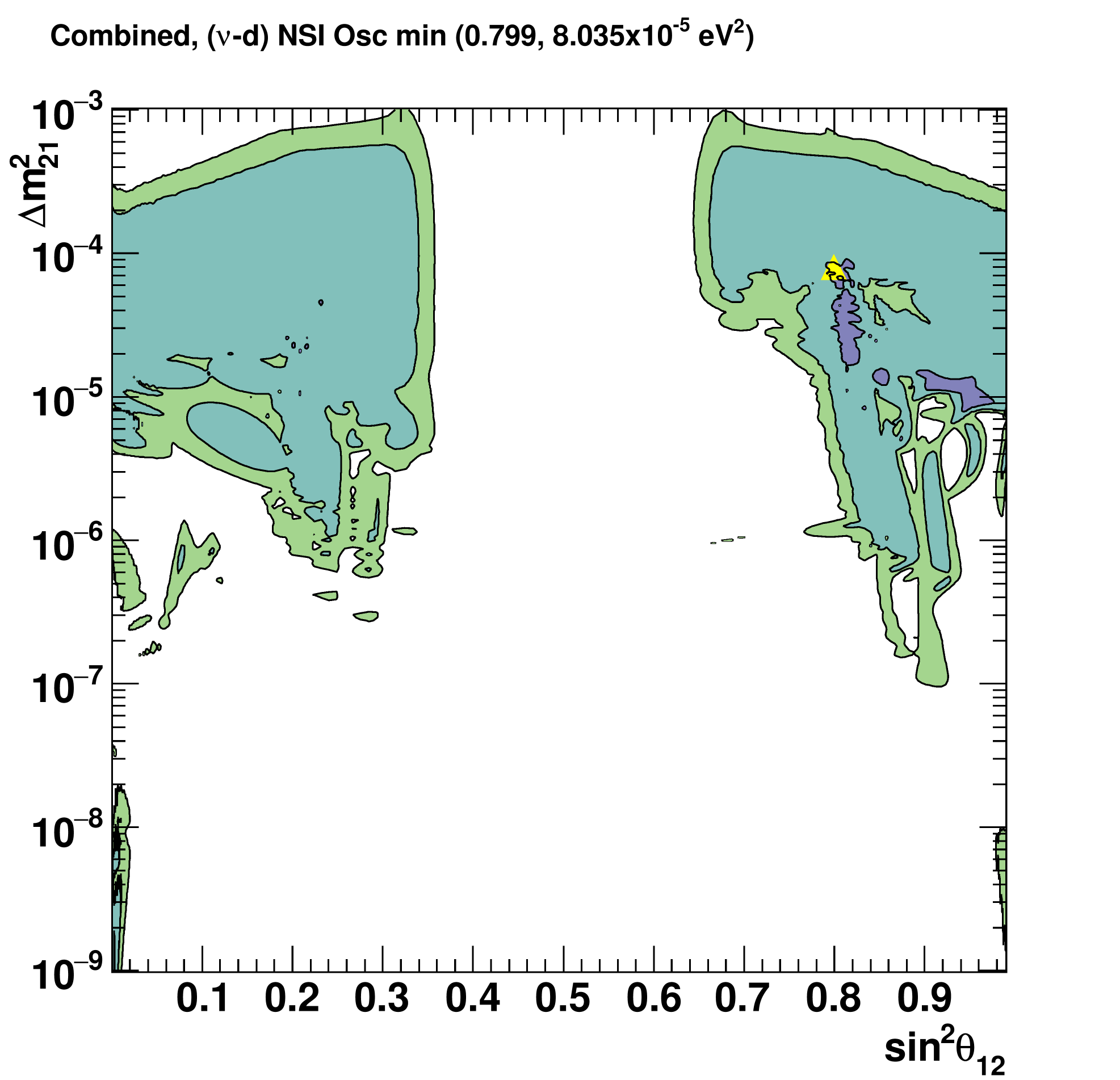} &
\includegraphics[width = 0.48\linewidth]{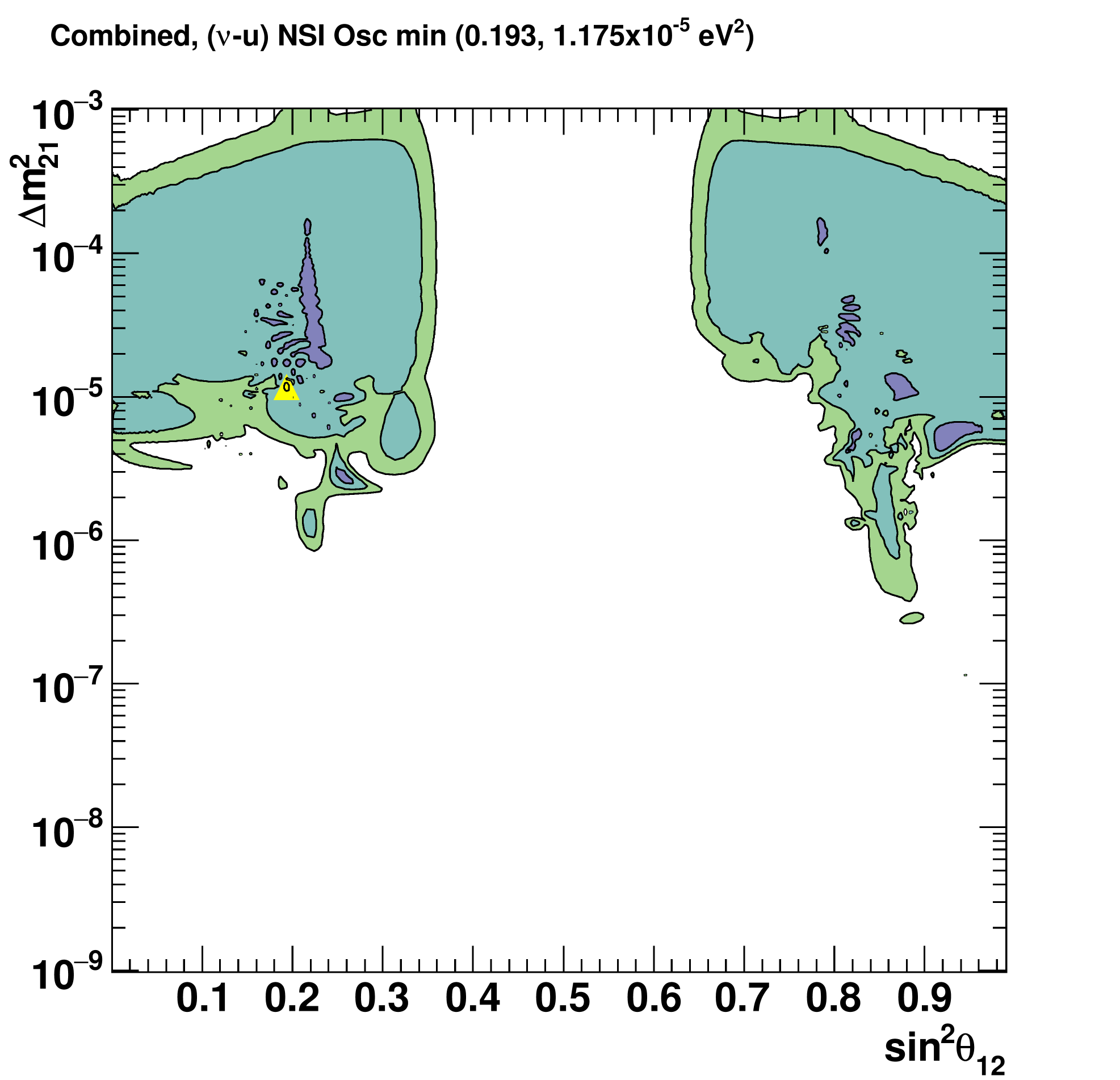} \\

\end{tabular}
\caption{The allowed regions given by the SK+SNO + Borexino $A_\textrm{DN}^\textrm{Be7}$ Combined fit for one (blue), two (teal), and three (green) $\sigma$ from two-dimensional parameter estimation ($\Delta\chi^2$ with 2 d.o.f).
The best-fit point is denoted by a yellow triangle and written in the title for each plot.
The left column corresponds to ($\nu$-$d$) NSI and the right column corresponds to ($\nu$-$u$) NSI.
The top row corresponds to the effective NSI parameter space ($\epsilon_{11}^f, \epsilon_{12}^f$), while the bottom row corresponds to the SI oscillation parameters ($\sin^2\theta_{12},\Delta m_{21}^2$).
The non-displayed parameters have been profiled over in each plot.}
\label{combined_results}
\end{figure*}

\begin{figure*}
\center
\begin{tabular}{c}
\includegraphics[width = 0.99\linewidth]{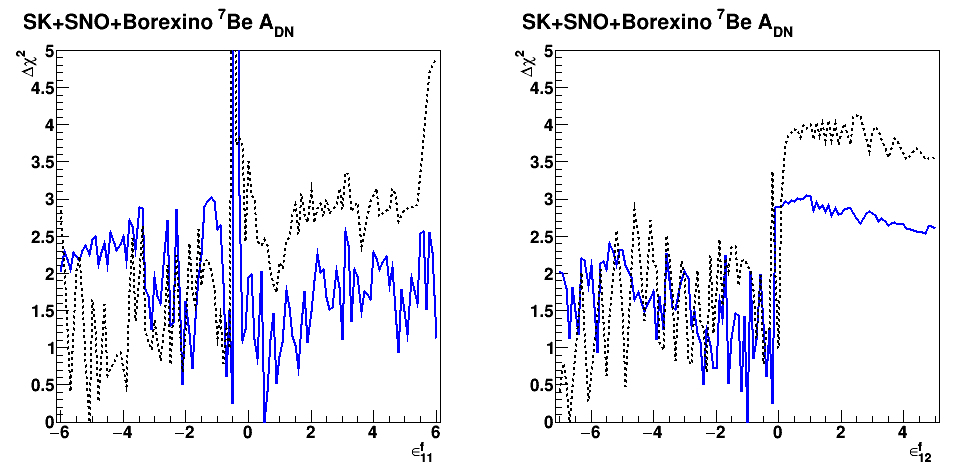}\\  
\end{tabular}
\caption{The one-dimensional $\Delta\chi^2$ limits from the Combined fit for the effective NSI parameters $\epsilon_{11}^f$ and $\epsilon_{12}^f$ for $f = u$ or $d$. 
The results for $\epsilon_{11}^f$ ($\epsilon_{12}^f$) are plotted in the left (right) column. 
The solid blue and black-dashed lines correspond to the results for ($\nu$-$u$) NSI and ($\nu$-$d$) NSI respectively.
All other parameters have been profiled over.}
\label{combined_1d_results}
\end{figure*}

After combining SK with SNO and Borexino $A_\textrm{DN}^\textrm{Be7}$ (the Combined fit) using the procedures described in Sec. \ref{sub:spec_pred} and Sec. \ref{sub:analysis_adn} respectively, 
the ($\nu$-$d$) NSI results for the $(\epsilon_{11}^{d},\epsilon_{12}^{d})$ parameter estimation contours have been determined.
These contours are plotted on the top-left panel of Fig. \ref{combined_results}, and the parameter estimation contours for $(\sin^{2}\theta_{12},\Delta m_{21}^{2})$  in Fig. \ref{combined_results} are plotted in the bottom left panel. 
The parameter estimation contours of the ($\nu$-$u$) NSI Combined fit is given on the right side of Fig. \ref{combined_results}. 
The $(\epsilon_{11}^{u},\epsilon_{12}^{u})$ allowed regions are given in the top right of the figure, and the $(\sin^{2}\theta_{12},\Delta m_{21}^{2})$ allowed regions are plotted in the lower left panel.
In all four plots, the best-fit point is denoted with a yellow triangle.
The corresponding one-dimensional $\Delta\chi^2$ limits for the effective NSI parameters are plotted in Fig \ref{combined_1d_results}. 
The results for $\epsilon_{11}^f$ ($\epsilon_{12}^f$) are plotted in the left (right). 
The solid blue and black-dashed lines correspond to the results for ($\nu$-$u$) NSI and ($\nu$-$d$) NSI respectively.
The Combined fit one-dimensional limits are similar to those from the SK-Only fit, though both $\epsilon_{11}^d$ and $\epsilon_{11}^u$ now disfavor the values around $\epsilon_{11}^f \simeq-0.4$ at $\Delta\chi^2>5$.
The range of $\epsilon_{11}^d > 0$ is still disfavored in the Combined fit at $\Delta\chi^2 > 1.7$.
For $\epsilon_{12}^u$, the Combined fit disfavors values larger than $-0.1$ at $\Delta\chi^2 > 2.5$, while $\epsilon_{12}^d$ above 0.2 is disfavored at $\Delta\chi^2 > 3.5$.

\begin{figure*}[]
\includegraphics[width = 0.99\linewidth]{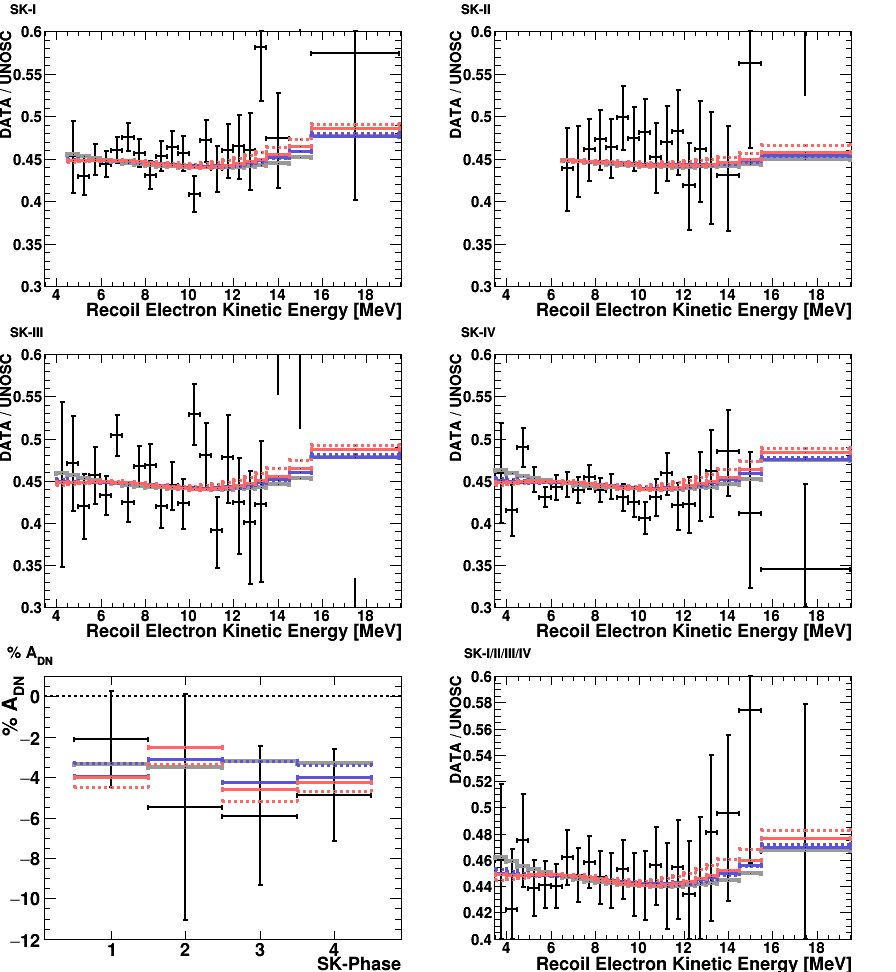}
\caption{The best-fit SK spectral and $A_\textrm{DN}$ predictions from the Combed fit to SK+SNO + Borexino $A_{\textrm{DN}}^{ \textrm{Be7} }$ (blue). 
The top four panels are the recoil electron spectra for each SK phase. 
The bottom left panel is the day-night asymmetry for each SK phase.
The bottom right panel is the combined SK-I/II/III/IV spectrum, which is provided for illustrative purposes.
The data (black lines) includes statistical and energy-uncorrelated errors added in quadrature.
The solid (dashed) lines correspond to NSI with up quarks (down quarks).
 The LMA best fit for SI (solid gray line) and the best-fit predictions for SK + Borexino $A_{\textrm{DN}}^{\textrm{Be7} }$ (red) are for plotted for comparison.}
\label{fig_comb_spec} 
\end{figure*}

\begin{figure*}[]
\includegraphics[width = 0.99\linewidth]{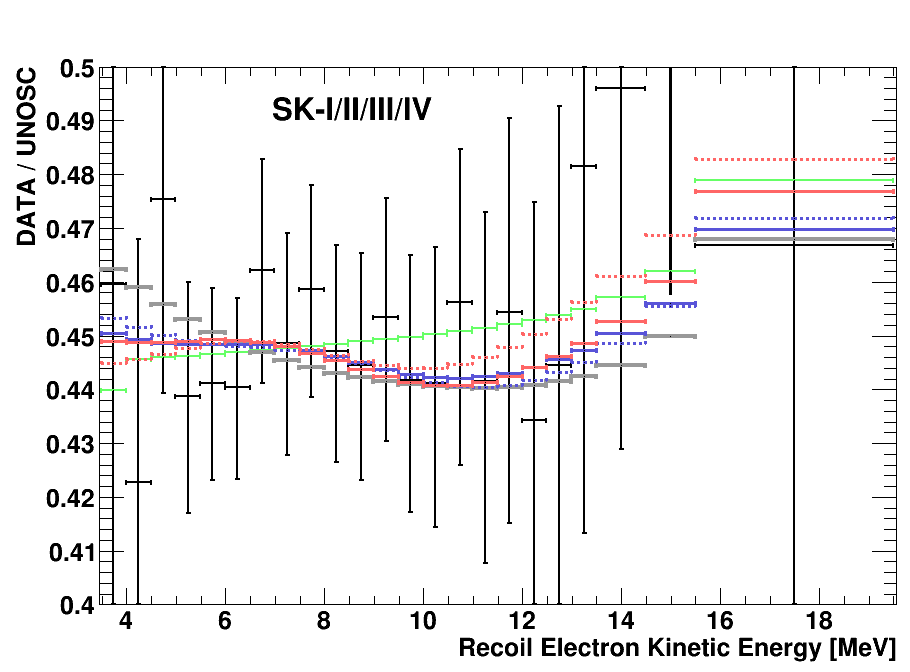}
\caption{The combined SK-I/II/III/IV spectrum from the SK+SNO+Borexino $A_{\textrm{DN}}^{\textrm{Be7} }$ Combined fit (blue). 
The solid (dashed) lines correspond to NSI with up quarks (down quarks).
 The LMA best fit for SI (solid gray line), the best-fit SK + Borexino $A_{\textrm{DN}}^{\textrm{Be7} }$ predictions (red lines), and the constant-value $P_{ee}$ = 0.317 prediction (green) are plotted for comparison.}
\label{sk1234_spectra_comb_nsi} 
\end{figure*}

The oscillation and NSI parameters for the best fit in each of the four results, along with the two results for the SI case, are summarized in Table \ref{nsi_results_table_large}.
The spectral and $A_\textrm{DN}$ predictions from the best-fit points from the Combined fit are plotted (blue lines) for each experimental phase of SK in Fig. \ref{fig_comb_spec} along with the SK+Borexino $A_\textrm{DN}^\textrm{Be7}$ best-fit results (red) from Fig. \ref{fig_skonly_spec}.
The combined SK-I/II/III/IV spectrum is provided for illustrative purposes in the bottom right panel and an enlarged version of the plot is provided in Fig. \ref{sk1234_spectra_comb_nsi}.

\subsection{Summary of Results}
For NSI between solar neutrinos and $d$-quarks ($u$-quarks), the SK-Only data prefers the best-fit NSI point to the best-fit SI point at a sigma-value of 
LLR $= 1.8\sigma$ $(1.6\sigma)$, while the combined fit shows a preference for NSI to SI at $1.9\sigma$ $(1.7\sigma)$.
When including the effective NSI parameters ($\epsilon_{11}^{f},\epsilon_{12}^{f}$) for $f = u$ or $d$, two additional degrees of freedom are added to the spectral fit, and the $\chi^2$ resulting from the maximum likelihood signal extraction method can be expected to improve by approximately the number of added degrees of freedom.
Because this improvement to the fit corresponds to an expected LLR of 1.0 (Eq.\ (\ref{llr_eq})) and an expected shift of $\sim$1.4$\sigma$ in the global minimum, the preference for NSI over SI is slight.
As one can see from the $(\sin^{2}\theta_{12},\Delta m_{21}^{2})$ contours, there is a non-negligible difference in the favored regions between the up and down quarks.
This difference predominantly comes from the different densities in the Sun, where $u$-quark density is higher than $d$-quarks, and thus have a stronger influence on the matter potential.
In the ($\nu$-$u$) NSI scenario, all $\Delta m_{21}^{2} \le 10^{-7} \textrm{ eV}^{2}$ are excluded by the combined fit, where as in the case of ($\nu$-$d$) NSI, islands of inclusion occur for very small $\Delta m_{21}^{2}$ when $\sin^{2}\theta_{12}$ is close to zero or one.
While the spectra and the day-night asymmetry for the best fits looks reasonable, the energy dependence of the SK day-night asymmetry and the energy spectra for the binned zenith results may provide more rejection power for NSI solutions that 
are currently viable in this analysis. 
 

\section{\label{sec:conclusion}Conclusions}
A fit to the SK recoil electron spectrum and day-night rate asymmetry data was performed while introducing effects on the neutrino wavefunction due to including neutrino-quark Non-Standard Interactions in the matter potential of the propagation Hamiltonian while independently varying the $(\theta_{12},\Delta m_{21}^{2})$ parameters and the effective NSI parameters $(\epsilon_{11}^{f},\epsilon_{12}^{f})$,  where $f$ = u-quarks or $f$ = d-quarks.  
The mixing angle $\sin^{2}\theta_{13} = 0.020$ was used for this analysis, and the NSI is assumed to couple neutrinos and quarks through a vector current interaction.
After calculating the $P_{ee}$ for $^{8}$B and $hep$ solar neutrinos during the day and night, the resulting recoil electron spectrum and $A_\textrm{DN}$ was compared to SK data, where the best fit with NSI is favored to the best fit with standard interactions (SI) at $1.8\sigma$ $(1.7\sigma)$ for d-quarks (u-quarks). After combining with the results from all three phases of SNO and Borexino's measurement of the $A_\textrm{DN}$ for $^{7}$Be solar neutrinos, the significance increases by $0.1\sigma$.

\section{Acknowledgements}
The authors gratefully acknowledge the cooperation of
the Kamioka Mining and Smelting Company. Super-K has
been built and operated from funds provided by the
Japanese Ministry of Education, Culture, Sports, Science
and Technology, the U.S. Department of Energy, and the
U.S. National Science Foundation. This work was partially
supported by the Research Foundation of Korea (BK21 and
KNRC), the Korean Ministry of Science and Technology,
the National Research Foundation of Korea (NRF-
20110024009), the National Science Foundation of
China (Grant No. 11235006), the European Union
H2020 RISE-GA641540-SKPLUS, and the National
Science Centre, Poland (2015/17/N/ST2/04064, 2015/18/
E/ST200758).

\bibliography{SK_Solar_NSI_apssamp} 

\end{document}